# Two-dimensional Magnetization-Transfer – CPMG MRI reveals tract-specific signatures in fixed rat spinal cord

Teresa Serradas Duarte and Noam Shemesh*

Champalimaud Neuroscience Programme, Champalimaud Centre for the Unknown, Lisbon Portugal

*Corresponding author:
Noam Shemesh
Champalimaud Neuroscience Programme, Champalimaud Centre for the Unknown, Av. Brasilia 1400-038, Lisbon, Portugal.
Phone number +351 210480000 ext #4467.
Email: Noam.Shemesh@neuro.fchampalimaud.org

Word count: ~8000 words
Number of Figures: 9 main figures (+7 supplementary figures)
Number of Tables: 2

# Abstract

Multiexponential T$_2$ (MET$_2$) Relaxometry and Magnetization Transfer (MT) are among the most promising MRI-derived techniques for white matter (WM) characterization. Both techniques are shown to have histologically correlated sensitivity to myelin, but these correlations are not fully understood. Furthermore, MET$_2$ and MT reports on different features of WM, thus being specific to different (patho)physiological states. Two-dimensional studies, such as those commonly used in NMR, have been rarely performed in this context. Here, off-resonance irradiation effects on MET$_2$ components were evaluated in fixed rat spinal cord white matter at 16.4T. These 2D MT-MET$_2$ experiments reveal that MT affects both short and long T$_2$ in a tract-specific fashion. The spatially distinct modulations enhanced contrast between microstructurally-distinct spinal cord tracts. Two hypotheses to explain these findings were proposed: either selective elimination of a short T$_2$ component through pre-saturation is combined with intercompartmental water exchange effects that occur on the timescale of irradiation; or other macromolecular species that exist within the tissue – other than myelin – such as neurofilaments, may be involved in the apparent microstructural segregation of the spinal cord (SC) from MET$_2$. Though further investigation is required to elucidate the underlying mechanism, this phenomenon adds a new dimension for WM characterization.

## Abbreviations:

B$_0$ – static magnetic field; BPF – bound pool fraction; CEST – Chemical Exchange Saturation Transfer; CNS – central nervous system; dCST – dorsal corticospinal tract; FC – fasiculus funeatis; FG – fusiculus gracilis; FOV – field-of-view; FWHM – full width half maximum; GM – gray matter; iLT – inverse Laplace transform; MTR$_{asym}$ – magnetization transfer ratio asymmetry; M$_{ref}$ – reference magnetization; M$_{sat}$ – saturated magnetization; MET$_2$ – Multiexponential T$_2$; MRI – magnetic resonance imaging; MS – multiple sclerosis; MSE – multi spin-echo; MT – magnetization transfer; MTR – magnetization transfer ratio; MWF – myelin water fraction; MW – myelin water; NMR – nuclear magnetic resonance; NNLS – non-negative least-square; NOE – Nuclear Overhauser Effect; osi – offset-saturation-induced; OW – other water; PBS – phosphate-buffered-saline; PCA – principal component analysis; PFA – paraformaldehyde; ReST – reticulospinal tract; RF – radio-frequency; ROI – region of interest; RST – rubrospinal tract; SC – spinal cord; SEM – standard error of the mean; SNR – signal to noise ration; STT – spinothalamic tract; T$_1$ – longitudinal relaxation time; T$_2$ – transverse relaxation time; TR – repetition time; TE – echo time; VST – vestibulospinal tract; WM –white matter.

# 1. Introduction

White matter (WM) microstructure integrity is essential for normal function of the central nervous system (CNS) [1]. Myelin content and composition are particularly important since they play a synergistic role with axon diameters in determining the conductance velocity along the axons [2–4]. While axon diameters can be characterized – to a certain extent – through advanced diffusion Magnetic Resonance Imaging (MRI) methods [5–8], multiexponential $T_2$ (MET$_2$) relaxometry and Magnetization Transfer (MT) are perhaps the most prevalent methods for studying myelin in tissues. Metrics derived from both MET$_2$ and MT have been shown to histologically correlate with myelin content, [9,10] even if in a different manner: the myelin water fraction derived from MET$_2$ is thought to better represent myelin content, while MT-based parameters are considered to better reflect myelin's chemical composition (although myelin is not the sole contributor to MT signal) [11]. Both MET$_2$ [9,12,13] and MT contrast [10,14,15] have been shown to be useful for investigating demyelinating diseases such as multiple sclerosis (MS) [16]. Additionally, MT techniques have been more recently combined with diffusion MR to extract microstructural features, namely, the g-ratio reflecting the ratio of inner and outer myelin sheath diameters [17].

MET$_2$ spectra obtained from Central Nervous System (CNS) white matter tissues typically exhibit two sizable components: a short $T_2$ component attributed to water trapped between the lipid bilayers of myelin; and a long $T_2$ component attributed to "other water", tentatively assigned to encompass contributions from both intra- and extra-axonal spaces [12,18–22] (in peripheral nervous system tissues, it is common to be able to observe separate peaks for myelin, intra- and extra-axonal spaces [23]). The two spectral peaks observed in CNS are typically identified through regularized inverse Laplace Transforms [24–26]. Mackay et al. mapped the short $T_2$ component fraction, termed by the authors "myelin water fraction" (MWF) as a first approach to measure myelin content [18]. MWF was shown to be histologically correlated with myelin content in several tissues both *in-vivo* and *ex-vivo*, including amphibian peripheral nerve [27], human brain [9] and rat spinal cord (SC) [28–30]. Also, MET$_2$'s applicability for studying demyelinating diseases such as MS has been demonstrated [9,12,13].

On the other hand, magnetization transfer imaging [31] shows signal attenuation when off-resonant RF irradiation is applied, typically in a range of ±100 kHz from the resonant frequency. Such attenuation is caused by the exchange of magnetic polarization between pools of nuclear spins. In MT, a two-pool model – mobile and bound proton pools – is frequently adopted. If several offset frequencies as saturation powers are used, quantitative information such as the bound pool fraction (BPF) can be estimated by fitting the data to a model [32,33]. The MT phenomenon is mediated by a wide range of mechanisms, including magnetization exchange (e.g. through dipolar- and J-coupling mechanisms, nuclear Overhauser effect (NOE – typically found in a range from -5 to -3 ppm), etc.), and chemical exchange of protons [32,34,35]. The latter can be investigated when focusing at a range of frequencies of 1-4 ppm – a technique termed chemical exchange saturation transfer (CEST), which has been receiving growing interest in the past years [36,37]. In WM, since myelin is one of the dominant source of macromolecules, MT-derived metrics provide an indirect, yet reproducible, measure of myelin content [38]. Histological validation of this hypothesis can be found in studies performed in a variety of tissues, including *e.g. ex-vivo* human brain [10], MS patients' spinal cords [39] and rodent models of demyelination/remyelination [40,41].

Although MWF and BPF (even MTR) derived from MET$_2$ and MT experiments, respectively, are histologically correlated with myelin content, these correlations are not straightforward. Importantly, while MT in tissue mainly emerges from dipolar couplings and chemical exchange, the origins of MET$_2$ are not yet fully understood. Clearly, populations characterized by different $T_2$s are required to invoke a multiexponential decay, but the identity of those pools and how they correlate with specific compartments in the tissue are not well understood. While at first glance, it would seem that MWF and the components undergoing MT in tissues should be similar, $MET_2$ and MT have been shown to be sensitive to different pathologies and/or physiological states [11,42,43]: for example, demyelination has been shown to induce decreases in MWF and MTR, but remyelination only affected MWF derived from $MET_2$, while MTR remained low [11]. In another study, MWF contrasted different spinal cord tracts that have similar myelin content but varying axon diameters, contrariwise to MTR [44]. Thus, MTR is thought to reflect differences in chemical composition of new myelin while MWF reflects some degree of intercompartmental water exchange [45]. These findings are in line with a recent study comparing MWF and BPF with myelin volume fraction in the mouse brain, calculated from a volumetric model of WM [46].

While studies comparing $MET_2$ and MT-derived contrasts provided vast input into the field, a combined, two-dimensional, experiment could potentially provide more detailed information than separate one-dimensional measurements [47]. Harrison et al. [48] performed a two-dimensional MT-multi spin-echo (MSE) (spectroscopic) NMR experiment at 1.5 Tesla, and reported the quantitative MT measurements to be identical for short and long $T_2$ components. Shortly thereafter, a hybrid MT-MSE NMR study with varying saturation pulse duration was conducted in bovine optic nerve, aiming to fit the data to a model comprising two tissue compartments, each with two proton pools, and incorporating exchange between them [49]. MT effects were then found to be mainly mediated by the short $T_2$ component, though water exchange was also found to influence the long $T_2$ component to a certain extent [49]. Similar results were obtained in an *in-vivo* human MR imaging study, reporting a larger MTR for the short $T_2$ component when the delay between the short-timescale MT modulation and acquisition was shorter than 200 ms [50].

Nearly all these studies observed how MT contrast varied at different TEs. However, if it is assumed that short $T_2$ components are more likely to comprise the BPF, a potentially instructive way to approach the experiment is from the opposite sense: to observe how MT modulates $T_2$ distributions extracted from $MET_2$. To our knowledge, whether magnetization transfer can actively affect the $T_2$ distribution in white matter has not been explored. In addition, the spatial distribution of such putative modulations has not been reported insofar.

Therefore, the goals of this study were: (1) to implement a two-dimensional hybrid MT-MSE technique to investigate how spectrally-resolved off-resonance irradiation modulates $MET_2$ components; (2) to assess whether the modulations are tract-specific. Upon performing such experiments in fixed rat spinal cord at 16.4T, we find intriguing modulations in white matter, which, we posit, could reflect exchange between water pools in the tissues or, alternatively, a more direct modulation by semisolid structures in the WM such as, hypothetically, neurofibrils. The contrast enhancements, as well as potential future directions, are discussed.

## 2. Materials and methods

This study was carried out in compliance with the recommendations of the directive 2010/63/EU of the European Parliament of the Council, authorized by the Champalimaud Centre for the Unknown's Animal Welfare Body, and approved by the national competent authority (Direcção Geral de Alimentação e Veterinária, DGAV).

*2.1.     Specimen preparation*

Male Long Evans rats ($N = 3$) were perfused transcardially with 4% paraformaldehyde (PFA) for extraction and isolation of the spinal cord. The samples were stored in 4% PFA for approximately 24 h to fix the tissue, after which they were transferred to a fresh Phosphate-Buffered-Saline (PBS) solution for 24 h. Before undergoing imaging (2 to 4 days after prefusion), cervical spinal cords were cut into thin sections of approximately 3 cm length and placed in a 5 mm NMR tube filled with Fluorinert (Sigma Aldrich, Lisbon, Portugal). A small plastic tube with diameter up to 1 mm was squeezed alongside the spinal cord to ensure the specimen remains stationary and does not move during scanning. A long plastic tube (with approximately 3 mm diameter) was positioned on top of the SC to prevent it from floating upward.

*2.2.     MRI experiments*

All experiments were performed on a 16.4 T Bruker Aeon scanner (Bruker, Karlsruhe, Germany), operating with Paravision 6.0.1 software, and interfaced with an Avance IIIHD console. The system is equipped with a micro5 probe and a gradient system capable of producing up to 3 T/m in all directions. A birdcage coil with inner diameter for 5 mm was used for transmitting and receiving at proton frequency. All experiments were conducted at 37°C.

To investigate multiexponential relaxation, a single-slice multiple spin echo (MSE) sequence was employed with the following parameters: FOV = (5.0 x 5.0) mm$^2$, matrix size = (80 x 80) leading to an in-plane resolution of (63 x 63) µm$^2$, slice thickness = 800 µm, echo time (TE) = {2.75:2.75:264.00} ms (96 increments), repetition time (TR) = 3000 ms, bandwidth (BW) = 100 kHz, number of averages (NA) = 4. These specific parameters were based on a preliminary $T_1$/$T_2$ study in the same spinal cords. A single slice in cervical portion of the spinal cord (C2-C7) was positioned in centre of the coil.

To achieve magnetization transfer, the MSE sequence was preceded by an MT module consisting of a gaussian-shaped RF pulse with a duration of 2 sec and peak power of 5 µT, and a gradient crusher prior to the MSE acquisition (**Supplementary Figure 1**). The selection of MT module parameters is explained in the **Supplementary Figure 2** and is consistent with other studies performed at 16.4 T [51]. MT offset frequencies were varied between {-50,50 kHz} in 1 kHz increments. The effect of any residual transverse magnetization following the off-resonance irradiation was removed by phase cycling the excitation pulse in two steps {0 2}, as was later confirmed by post-hoc analysis (see below). The total of 102 experiments took ~30 h per spinal cord in total.

*2.3.   Image Analysis*

Image processing and analysis were performed in MATLAB (The Mathworks, Natick, MA). Magnitude data was converted to real data using Eichner's method [52]. Subsequently data was denoised using Veraart's principal component analysis (PCA) algorithm based on random matrix theory and Marchenko-Pastur distributions of noise eigenvalues [53]. The echo times were used as the redundant dimension, and the denoising was repeated for each MT offset frequency. Finally, a last Gibbs unringing step based on the algorithm presented by Kellner et al. [54] was employed. Illustrative images of intermediary steps in the pre-processing pipeline are shown in **Supplementary Figure 3**.

Region of interest (ROIs) analysis was performed on two ROIs representing WM and GM. Large portions in the lateral WM and central part of GM, respectively, were taken to ensure sufficient signal to detect even small changes. In addition, to assess tract-specific information, ROIs were carefully chosen for the seven different tracts in WM [44,55–58], including dorsal Corticospinal tract (dCST), Fasiculus Cuneatis (FC), Fusiculus Gracilis (FG), Reticulospinal tract (ReST), Rubrospinal tract (RST), Spinothalamic tract (STT) and Vestibulospinal tract (VST). In all cases, the average of the ROI signal intensity was used for further analysis.

*2.3.1. Magnetization Transfer analysis*

Magnetization Transfer Ratio (MTR) [59] and MT asymmetry (MTR$_{asym}$) [60,61] at every TE were evaluated according to

$$MTR = \frac{M_{sat}(\omega)}{M_{ref}} \tag{1}$$

$$MTR_{asym} = \frac{M_{sat}(-\omega) - M_{sat}(\omega)}{M_{ref}} \tag{2}$$

where M$_{sat}$(ω) is the signal intensity at an offset frequency ω, and M$_{ref}$ the intensity of the reference image. MTR and MTR$_{asym}$ maps of a single offset frequency irradiation (and single TE) were also generated for the frequency with maximum $|MTR_{asym}(\omega)|$. In each ROI, the Mean±SEM among SCs are reported.

*2.3.2. Multiexponential T$_2$ analysis*

For the MET$_2$ relaxometry analysis, the multi-exponential signal decay over time *t* can be written as:

$$S(t) = \sum_i A(T_{2,i}) \cdot e^{-\frac{t}{T_{2,i}}} + \varepsilon \qquad (3)$$

where $T_{2,i}$ and $A(T_{2,i})$ correspond respectively to the relaxation times (T$_2$s) and the T$_2$s distribution fractions, and $\varepsilon$ represents the noise. The T$_2$ distribution can be calculated by a Laplace inversion [18,62], which decomposes the signal into an arbitrary number of exponentials, allowing the representation of T$_2$ distributions with varying number of components. A non-negative least square (NNLS) algorithm can be used for this purpose [18–20,24,62], with all the cautions of ill-posedness. In this study, we used a regularized non-negative least square (NNLS) algorithm [24–26] regularized with minimum curvature constraints as in [44]. Extended phase graph analysis was performed to account for B$_1^+$ inhomogeneities and ensuing stimulated echoes [63]. The code used for the Laplace inversion is part of the REMMI analysis tool and can be found online in http://www.vuiis.vanderbilt.edu/~doesmd/MERA/MERA_Toolbox.html.

The regularization weighting was set to α = 0.05 and the number of points in the T$_2$ space were 200, spaced from 2 to 265 ms. The NNLS procedure was conducted both for the mean signal decay in each ROI and voxel-by-voxel for all 102 sets of images (101 different offset frequency images + 1 reference). The fit error was calculated as the sum of the squared residuals. The SNR was also calculated using the REMMI analysis tool box, and it is given by the sum of the T$_2$ spectrum divided by the standard deviation of the residuals. In order to evaluate the effect of noise on the calculation of T$_2$ spectra, increasing levels of gaussian noise was added to the data from one representative high SNR ROI in the WM, and the respective T$_2$ distributions were analysed. (**Supplementary Figure 4**). A minimum mean SNR (averaged in all SC voxels) of 500 was set for exclusion of samples.

*2.3.2.1. Analysis of $T_2$ distribution peaks*

The distributions were divided into $T_2$ components (short/long $T_2$) according to the characteristics of its peaks. Two parameters were extracted per peak/component: *i)* fraction of components and *ii)* $T_2$ of peak. The short $T_2$ component commonly referred to as *myelin water*, was computed for each spectrum from the peak with $T_2$<23.5 ms. The long $T_2$ component (or other water) exhibited a peak between 23.5 and 65.0 ms. For each offset frequency irradiation, two parametric maps of each component were generated from the voxelwise inverse Laplace inversions – myelin water fraction (MWF) *vs* other water fraction (OWF) maps and $T_2$ short *vs* $T_2$ long peak maps.

*2.3.3. Hybrid MT-MSME: $MET_2$ changes over offset frequency*

Finally, to investigate if MT irradiation modulates $MET_2$ relaxometry, the variation of $T_2$ distributions over offset frequency was explored in each ROI. 3D contour maps of $T_2$ distributions were plotted against the offset frequency. Then, each one of the four extracted component-based parameters were plotted against the irradiation frequency. Both absolute and relative changes (normalized to the reference image) of each parameter were calculated.

To facilitate the visualization of differential spatial effects of such spectra, maps of offset-saturation-induced changes on $MET_2$ parameters were derived. To do so, four spectral windows with a 15 kHz range were selected: Window 1= {-45,-30} kHz; Window 2 = {-20,-5} kHz; Window 3 = {5,20} kHz; Window 4 = {30,45} kHz. Windows 1 and 4, and 2 and 3, respectively, are symmetric with respect to the water resonance. In order to obtain single-image maps, the integral of each window was taken per coefficient per voxel, resulting in parametric offset-saturation-induced-$MET_2$ window maps. Voxel-by-voxel interpolation of parametric changes over frequency was conducted prior to the integration. Lastly, two difference maps were calculated: one with two symmetrical spectral windows (Window 3 - Window 2); and another map with the difference between a window far from the resonance frequency representing "unsaturated" magnetization (Window 4) and a spectral window with an accentuated MT modulation (Window 3).

# 3. Results

Raw (MT-)MSE data from a representative spinal cord are shown in **Figure 1A** and **Figure 1B**, enabling the assessment of image quality. **Figure 1A** displays representative images after denoising for different echo times, and **Figure 1B** shows the entire 2D decay profile for the white matter ROI. In both dimensions, the signal is well-behaved, decaying non-monoexponentially in the TE domain and evidencing z-spectra shapes in the irradiation offset domain. The SNR (averaged among all voxels in the SC) of each spinal cord (SC#1, #2 and #3) in the first echo of reference image (without MT) was 506, 537 and 358, respectively. SC#3 did not reached the minimum SNR criterion and for that reason was excluded from further analysis. The difference between non-denoised and denoised images showed only noise and no structure, indicating a good performance of the denoising procedure (**Supplementary Figure 3**).

*3.1. Magnetization Transfer*

To select specific parameters for the 2D experiments, MT z-spectra and $MTR_{asym}$ (Mean±SEM among SCs) were assessed at specific TEs for different ROIs (**Figure 2A**). Both MTR and $MTR_{asym}$ plots are in line with what has been observed in white matter, where a shift in the MT symmetry with respect to water is predicted due to the contribution of the semisolid pool [60,64].

As expected, MT effects are larger in WM compared to GM ROIs (**Figure 2B**). However, tract-specific information is much less apparent: both MTR and $MTR_{asym}$ maps for ω = -13 kHz, corresponding to the largest observed MT effect in WM (**Figure 2C**), evidences only minor contrast among the different WM tracts. MTR and $MTR_{asym}$ plots of the respective tract ROIs supports this observation, showing very small differences in MT for these tracts (**Figure 2D**). **Figure 2E** compares z-Spectra and $MTR_{asym}$ plots of the WM ROI at different echo times. Greater MT effects were observed for shorter echo times.

### 3.2. MET$_2$ relaxometry

Prior to the 2D experiment, specific MET$_2$ experiments were examined. In white matter, a non-monoexponential MSE signal decay was observed, giving rise to a distribution of T$_2$s via the inverse Laplace transform (**Figure 3**). In all spinal cords, two peaks were detected in WM around 10-20 ms and 25-50 ms, while only one peak around 25-50 ms was found in GM. In white matter, the short T$_2$ component represented a fraction of (Mean±SEM among SCs) 0.43±0.01 of the total T$_2$ distribution and was attributed to "myelin water" (MW) [44]. The long T$_2$ component, representing the reciprocal fraction of 0.57±0.01, was considered to correspond to the "other (or intra-/extra-axonal) water" (OW) [44]. Parameters extracted from the inverse Laplace transform (iLT) in experiments without any MT pulses are summarized in **Table 1**; results from one representative spinal cord are shown in **Figure 3**. Two parameters were extracted per T$_2$ component, its fraction and the T$_2$ of peak.

*Table 1 – T$_2$ distribution parameters from inverse Laplace analysis in the WM of all SCs*

| Slice position | T$_2$ short | | T$_2$ long | |
|---|---|---|---|---|
| | T$_2$ of Peak (ms) | fraction | T$_2$ of Peak (ms) | fraction |
| SC#1 (~C2) | 18.60 | 0.44 | 41.18 | 0.56 |
| SC#2 (~C4) | 16.37 | 0.42 | 37.07 | 0.58 |
| Mean±SEM | 17.48±1.11 | 0.43±0.01 | 39.12±2.05 | 0.57±0.01 |

Voxelwise parametric maps of the short and long T$_2$ components are shown in **Figure 3B**. The extracted parameters were found to contrast different WM tracts well. **Table 2** registers the Mean±SEM (among SCs) T$_2$s and areas of short and long T$_2$ components of different ROIs (WM, GM and seven different WM tracts). dCST was the ROI with highest variability and biggest number of outliers among ROIs and SCs.

*Table 2 – Mean±SEM of T2 distribution parameters of different ROIs*

| ROIs | $T_2$ short | | $T_2$ long | |
|---|---|---|---|---|
| | $T_2$ of Peak (ms) | Area (fraction) | $T_2$ of Peak (ms) | Area (fraction) |
| WM | 17.5±1.1 | 0.43±0.01 | 39.1±2.1 | 0.57±0.01 |
| GM | N/A | 0 | 30.7±1.8 | 1 |
| dCST | 16.5±0.1 | 0.21±0.002 | 28.2±0.3 | 0.79±0.002 |
| FC | 16.0±0.5 | 0.45±0.007 | 42.0±1.4 | 0.55±0.007 |
| FG | 14.0±0.9 | 0.29±0.04 | 31.1±2.1 | 0.71±0.04 |
| ReST | 17.2±0.4 | 0.42±0.01 | 39.0±0.1 | 0.58±0.01 |
| RST | 16.8±2.0 | 0.37±0.03 | 36.2±3.6 | 0.63±0.03 |
| STT | 17.4±0.3 | 0.45±0.04 | 37.6±2.1 | 0.55±0.04 |
| VST | 18.5±1.15 | 0.46±0.02 | 44.2±1.9 | 0.54±0.02 |

WM – white matter; GM – grey matter; dCST – dorsal corticospinal tract; FC – Fasiculus Gracilis; FC – Fusiculus Cuneatis; ReST – Reticulospinal tract; RST – Rubrospinal tract; VST – Vestibulospinal tract

### 3.3. Offset-Saturation-Induced (osi-) Shifts in MET$_2$

To assess whether $T_2$ distributions are modulated by MT saturation, iLT analysis was performed for every MT offset frequency. **Figure 4** shows the results for a representative SC. In the WM ROI, a clear modulation of $T_2$s is observed (**Figure 4A**), which we term "Offset-Saturation-Induced" (osi-) MET$_2$ shifts. The GM ROI also exhibits modulations in the (single-peaked) spectra. By visual inspection of the contour plots of $T_2$ distributions over frequency in different ROIs, these modulations appear to have tract-specific characteristics (**Figure 4B**): for example, though all distributions seem to be shifting towards lower values as the saturation frequencies become closer to zero (approximately symmetrically), the dCST shows a larger modulation compared to, *e.g.* VST.

A more quantitative analysis of the osi-shifts was conducted by examining the irradiation-induced modulations on the two parameters that characterize the short and long $T_2$ components in the distribution. **Supplementary Figure 5** shows that the variability in spinal cord position or preparation does not greatly impact the reproducibility of osi-shifts. Hence, **Figure 5** shows the Mean±SEM osi-shifts among all SCs for the ROI defined in **Figure 5A**. **Figure 5B** and **Figure 5C** correspond to absolute and relative parameter changes in a large WM ROI, respectively. The maximum variation in mean short $T_2$ was ~7 ms, from 10.0±0.5 ms to 16.65±0.7 ms, with its fraction varying from 0.26±0. 0.001 to 0.41±0.001. The mean long $T_2$ component in WM ROI varied by ~4 ms, from 33.5±1.5 ms to 37.3±1.9 ms, with its mean fraction increasing from 0.58±0.001 to 0.74±0.001. As predicted, both the absolute and relative variation of the short $T_2$ component with MT modulation is larger compared to the variation of the long $T_2$ component. Both parameters exhibited a similar trend (**Figure 5C**).

To study whether osi-shifts may be tract-specific, the same spectral signatures were compared among different WM tracts (c.f. **Figure 6A** for ROI definitions). Different tracts reveal both variability on the absolute initial $T_2$s (particularly on the long $T_2$ component, see **Figure 6B** and **Figure 6C**) and on the amplitude of osi-shift effect (**Figure 6B** and **Figure 6C**). Interestingly, dCST and VST, as well as FC, that are among the anatomically most different WM tracts *vis-à-vis* their average diameters, presented with the biggest variation among tracts, both on the absolute value of parameters regardless of MT modulations and on the amplitude of osi-effects after irradiation. Please note that the high variations particularly noticeable in the dCST tract are induced by noise inherent to the inverse Laplace transform.

The specificity of some osi-shift parameters to specific tracts raise the question whether osi-shifts are spatially coherent. To investigate this, parametric maps of osi-shifts in specific spectral windows were plotted (**Figure 7**). The selection of spectral windows is illustrated in **Figure 7A**, where four symmetric "two-by-two" windows (window 1 is symmetric to window 4 while the window 3 is symmetric to window 2) were defined based on similar osi-shift characteristics. Maps of each parameter in each specific spectral window are shown in **Figure 7B**. The maps are symmetric for all spectral windows, both for long- and short-component osi-shift parameters, suggesting that the osi-shifts are spatially coherent. Interestingly, nearly all maps of spectral windows 2 and 3, which are those with larger MT saturation, exhibit higher contrast among different WM tracts. This phenomenon is particularly evident in short $T_2$ fraction (or MWF) osi-shift maps (**Figure 7**). Then, difference maps were calculated between the different spectral windows. The difference between symmetric saturated spectral windows did not exhibit any specific contrast within the white matter, suggesting indeed that the maps are symmetric with respect to 0 frequency **Figure 7D**). By contrast, the difference between spectral windows with and without saturation exhibited a distinct contrast within the white matter (**Figure 7C**).

To investigate whether these spectral window maps can be used to enhance the contrast to specific WM tracts, three profile lines were placed along different tracts in Spectral Window, MWF and MTR maps (**Figure 8A**). The variation of the signal along these lines was computed and compared for the three different techniques (**Figure 8B**). Finally, the Mean±STD of the signal intensity within each of the seven WM tracts ROI was plotted to have a more quantitative analysis of the same quantity (**Figure 8C**). The Spectral Window maps appear to be able to enhance the contrast between different WM tracts (see e.g. ReST *vs* VST), although more SC and further statistic tests beyond the scope of this work is required to confirm our hypothesis.

## 4. Discussion

MRI's potential for studying healthy and diseased white matter is one of its hallmarks. Obtaining more detailed and specific information on myelin is perhaps one of the greatest challenges in contemporary white matter MR imaging [12,17,46]. Although it is well-stablished that both MET$_2$ and MT metrics are histologically correlated with myelin content [9,10], a better understanding of each method's source of contrast is required for accurately extract quantitative information about the tissue from these metrics, such as myelin fraction, integrity, or its composition. Previous studies have found that mainly the short T$_2$ component undergoes MT, although the longer T$_2$ components may also be influenced – directly or indirectly – by the irradiation [49,50]. However, whether the T$_2$ distribution in its entirety could be modulated by MT irradiation, its dependence on the full irradiation frequency, and the spatial distribution of these effects was not investigated, to our knowledge.

(1) Here, we used a 2D MT-MET$_2$ MRI approach to investigate modulations of T$_2$ distributions in fixed rat spinal cords at 16.4 T. Several findings can be noted from our study, and will be discussed in-depth below: T$_2$ components extracted from the analysis in this study, even without irradiation, were characterized by relatively long MW T$_2$s (MW ~10-20 ms; Mean±SEM = 17.5±1.1 ms) compared to *other ex-vivo* MET$_2$ studies. MWF was also uniformly higher. OW T$_2$s, on the other hand, were comparable or even slightly shorter (OW ~25-50 ms; Mean±SEM = 39.1±2.1 ms) [28,44];

(2) MT irradiation modulates the T$_2$ spectra and shifts both short and long T$_2$ components (although the effect is larger in short T$_2$ – MW – component);

(3) These osi-shift signatures appear to have some specificity in terms of contrasting different spinal cord tracts.

*Long T$_2$s in ex-vivo spinal cords at ultrahigh field*. In fixed spinal cord white matter, we observed peak T$_2$s of MW and OW at ~10-20 ms and ~25-50 ms, respectively. These values may be judged to be high considering the ultrahigh 16.4 T magnetic field. Previous studies by Kozlowski et al. in *ex-vivo* spinal cord at 7 T reported T$_2$s of MW <20 ms and OW <50 ms [28] and similarly, Dula et al. observed T$_2$s of MW ~10-14 ms and OW ~43-61 ms [44]. *In-vivo*, T$_2$s of the spinal cord at 7 T were reported at MW <20 ms and OW >30 and <100 ms [28]. Oakden et al. reported MW at 3-39 ms (healthy and injured) and healthy OW at 40-77 ms [29]. We attribute the slight long MW T$_2$ measured in this study to several mutually-reinforcing factors. First, the SNR was *a priori* high, which will assist in an accurate determination of T$_2$. Second, the interecho spacing was short (2.75 ms), which provided a high-fidelity curve for analysis. Third, the MP-PCA denoising [53] significantly improves SNR without blurring, and "revives" signal at the tails of the decay. Thus, denoised data can be expected to have the more accurate and longer T$_2$ compared to their non-denoised counterparts. Fourth, the spatial resolution in this study was higher than those previously reported, thereby reducing partial volume effects. Finally, and probably most importantly, unlike Dula et al., [44] and Kozlowski et al., [28], our experiments were conducted a temperature of 37 °C. A-priori, the higher temperature will increase T$_2$ values due to increased molecular rotational degrees of freedom, but it also will increase exchange which can explain why the longer T$_2$ component was not much higher. Indeed, when we repeated some of these experiments at lower temperature, the T$_2$ spectral trends are consistent with ex-vivo spinal cord literature (data not shown), which was invariably performed at room temperature rather than 37 °C.

*Modulations of T$_2$ spectra by MT irradiation.* The 2D MT-MET$_2$ pulse sequence in this study was designed to pre-saturate one component and observe the ensuing T$_2$ spectrum. If two non-interacting components with distinct $T_2$s existed, and only one of those species (assumingly the short $T_2$ component) could undergo MT, then the experiment would have yielded a $T_2$ spectrum where only one component's area is modulated. That is, upon "erasing" magnetization from one species, it's magnitude in the $T_2$ spectrum would be lowered (as there are less spins to contribute to $T_2$ relaxation), while the other species' spectrum would not be affected.

Before analysing MT irradiation effects in $MET_2$ components, it is worth considering the irradiation effects at different echo times (**Figure 2**). The frequency range employed was quite large, but the quite selective irradiation employed here may also affect CEST contrast [37,65] in the relevant frequency range of ~ 0.7-2.8 kHz (1-4 ppm). Our experiments only contain 3 acquisitions close to this range (1, 2, 3 kHz), which precludes a detailed analysis of CEST-related phenomena. A similar argument applies to NOE, where the frequency range of -5 to -3 ppm [66] is poorly characterized in our experiment. The characteristic shape observed in MTR and $MTR_{asym}$ plots at different echo times thus appears to be related to a shift with respect to the water frequency induced by the presence of the macromolecular pool itself [60,64].

When focusing at the effects of MT modulation on the parameters extracted from the $T_2$ spectra analysis, our findings show that both MW and OW are affected by irradiation, though to different extents. Modulations of MWF – the short $T_2$ component – were expected and discussed in the past: combined MT-$MET_2$ NMR/MRI studies in WM [48–50,67] reported significant larger MTR for the short $T_2$ component when compared to the long $T_2$, if the timescale of the MT modulation is shorter than 200 ms. Such results support that MT effects are mainly mediated through the short $T_2$ component, although for longer saturation time experiments an average MT effect is observed by providing time for the components to exchange. These studies have also shown MT effects on OW, and indeed, in our study OW was also found to be modulated by the MT (**Figure 5**), although an overall greater effect on MW (short $T_2$) was observed.

It might be worthy to notice that these studies have been performed in different types of myelinated tissues, e.g. in human brain [42] and in peripheral nerve [67]. The brain white matter consists of smaller axons while peripheral nerve typically has larger axons and more myelin. In our case, we studied an "intermediate"-level tissue, the rat spinal cord, which typically has larger axons than the brain but smaller than the peripheral nerve. In general, a shift in $T_2$ distributions were observed also in those previous studies (though not thoroughly discussed), but the magnitude and even direction of shifts appear to vary substantially. In particular, both Vavasour and Does studies show an increase of the shortest $T_2$ while here a decrease of all $T_2$s is observed. The common feature in all studies is that the strongest effect was shown to mainly involve the shortest myelin-related peak. Further comparisons of osi-shifts for different types of tissues would be potentially very informative, but requires much more similar acquisition conditions (field strength, irradiation pulse, *etc.*).

Note also that SNR loss with saturation is not a likely source for osi-shifts. **Supplementary Figure 6** hows an additional spectrum acquired with increased NA closer to the water resonance. Increasing the SNR of images with high MT modulation did not change the previously-observed osi-shift trends, suggesting that SNR was not a dominant constraint in this study. Note that this is a complement of **Supplementary Figure 4** (rather than a replacement of the same) for a better characterization of the effects of noise on the $T_2$ distributions.

At least two distinct plausible explanations for these observations can be invoked: (1) the OW contain axonal and extra-axonal water, which will interact with the innermost and outermost layers of myelin, respectively, and the long offset saturation may create an efficient labelling mechanism for such water; (2) exchange between the MW and OW may be significant enough that saturation of one component then imprints on the other. A third, potential explanation would be that OW interacts with macromolecules (other than myelin) in the axonal space – for example, neurofibrils and microtubules, which in turn may also induce direct MT effects. Neurofibril staining is actually a marker for intra-axonal space [57], and neurofibrils are quite densely packed; in this case, the neurofibril density may be proportional to the axon diameter, which can explain both the contrast in MET$_2$ images (through diffusion-mediated decoherence) as well as the MT influence of the longer T$_2$ component.

*Tract-specific contrast of osi-shift signatures.* One key observation is that osi-shifts are tract specific. Moreover, the ability of osi-shifts from a combined MT-MSE study to contrast different tracts appears to be higher than any of these techniques alone (**Figure 8**).

The biggest differences in osi-shifts were noted for tracts which are histologically known to have the largest disparity in micro-anatomical features, and particularly, the axon size distribution (**Figure 9**). The dCST has the smallest axons in the SC, with average axon diameters around ~1.4 µm, while FC and VST have mean axon diameters of ~3.4 µm as measured from histology [57]. The ReST has been characterized by having the largest diameters, with average axon size of ~4 µm, but has lower axonal density compared to dCST and FC, or VST [44,56,57]. Importantly, the myelin fraction is quite similar, varying by less than 10% among all tracts, between 0.35-0.39 [57] (**Figure 9A**). Indeed, the osi-shifts showed dCST and FC, or VST, to be the most disparate in terms of the osi shift amplitudes. This observation can provide additional insight in the context of potential explanations for the observation of osi-shifts in MW and OW. Since the myelin content is not very different between the tracts, yet the osi-shifts seem to vary correspondingly with axon sizes, a surface-to-volume effect could be invoked as a potential mechanism. In particular, assuming that exchange rate (MW → OW) would be similar among tracts, a large MT effect for the smallest tracts could indicate either that exchange is greatest (the myelin thickness would be the smallest in the smaller tracts) or, alternatively, that restricted diffusion enables more cross-relaxation at the boundaries and macromolecules within the intra-axonal space. The latter hypothesis would be in line with the involvement of neurofibrillary macromolecules in $T_2$ contrast.

Alternatively, these modulations may reflect intercompartmental water exchange, which may cause an underestimation of short $T_2$ component area and $T_2$ values. This hypothesis has been proposed before to explain contrast in MWF maps in [44,45]. Variations in MW and OW over the saturation frequency (after MT modulation) would then be a consequence of affecting myelin through presaturation, which would then be transferred to the OW compartment by water exchange. Further studies would be required to validate or exclude any of the proposed explanations.

### 4.1. Technical considerations

The spinal cord was selected as a model of WM due to its diverse microanatomy: it is composed by a variety of tracts with different mean axon diameters and myelin thickness despite having similar myelin densities [44,56,57]. Additionally, it is a structure with mostly aligned fibers [56,57] which makes it easier to study as $T_2$ anisotropy could be expected to be small [68]. Notwithstanding, several factors must be considered to ensure the reliability of the results.

*$B_1$ inhomogeneities.* The transmit field inhomogeneity can significantly influence the determination of $T_2$ estimation quality [28]. In this study, a 5-mm birdcage coil enclosing the entire SC was used. This setup incurs minimal $B_1$ inhomogeneities as confirmed by using a double angle method experiment [69]: indeed, the $B_1$ map (**Supplementary Figure 7**) shows that there are little deviations from the prescribed flip angles. Moreover, $B_1$ inhomogeneities and ensuing stimulated echoes were reduced by performing extended phase graph analysis as described before in [63].

*Diffusion attenuation in a CPMG.* In a CPMG-like sequence, diffusion weighting induced by imaging gradients may contribute some attenuation [70] which, if unaccounted for, can bias $T_2$ estimation. However, the short delays in our pulse sequence (2.75 ms) render diffusion weighting small. Hence, we can exclude diffusion-based inaccuracies as the dominant source of the observed shifts in $T_2$ spectra.

*SNR, interecho spacing, minimum first echo time and spatial resolution.* All these are important factors in quantifying $T_2$ spectra, especially given the ill-posedness of the iLT. The SNR of the experiments in general was very high (>500), both due to the ultrahigh field strength and due to the Marchenko-Pastur PCA denoising, which can remove noise without compromising image resolution [53]. Moreover, the SNR change over the probed frequency spectrum could not explain the osi-shifts (see **Supplementary Figure 4**), although it must be recognised that the effect of direct saturation when approaching the water frequency might affect $T_2$ distributions. therefore, less attention was given to frequencies ~5 kHz around water resonance, which contain the strongest direct saturation effects. The interecho spacing employed here was also probably one of the shortest reported for MRI-based multi-echo sampling in the context of myelin water (2.75 ms). This increases the accuracy of the $T_2$ spectrum estimation. Similarly, our minimum echo time of 2.75 ms was short, increasing the quality of $T_2$ spectra. Finally, high spatial resolution was used, which *(a)* reduces internal-gradient-induced diffusion [71] and *(b)* reduces partial volume effects, both of which then provide a better estimate of $T_2$ [28].

*Comparison of MTR, MT$_{asym}$ and MET$_2$.* It is interesting to compare the different contrasts obtained from each method, prior to discussion of the 2D approach. The two parametric maps extracted from conventional MET$_2$ spectra (**Figure 3B**) were compared with MTR and MTR$_{asym}$ maps from the same spinal cord (**Figure 2C**). While T$_2$-derived parametric maps contrast between the SC's different tracts, particularly T$_2$ and area (MWF) maps, MTR and MTR$_{asym}$ maps were considerably more spatially homogeneous in WM (only slight contrast between FG *vs* FC/dCST can be depicted in MTR maps). This observation is in agreement with the findings by Dula et al. [44], who reported MWF to vary with microanatomy whereas MT characteristics remain constant (as mentioned before, the authors proposed intercompartmental water exchange to justify such difference, an hypothesis supported by simulations [45]). Such remarks reinforce the idea that MT is more sensitive to the biochemical composition of tissue while T$_2$s appear to contrast different physical compartments [11].

### 4.2. *Potential application of osi-shifts*

Finding osi-signatures to be tract-specific can potentially represent an opportunity: we hypothesized that the osi-shifts could be used to enhance contrast to specific tracts by mapping the integral of a specific peak or spectral window in the osi spectra. **Figure 8** shows that indeed tracts with different properties are contrasted using the spectral window maps, although further comparisons would be needed not only to test the gains of this method versus standard MWF mapping but also the exact source of such contrasts.

It is noteworthy to realize that several factors would have to be considered if looking beyond excised tissues into *in-vivo* applications. Though the conduction of experiments at 37 ºC already represents an environment closer to that of *in-vivo* experiments, chemical fixation is known to reduce $T_2$s, although most of these effects are reversed after washout [72,73]. Indeed, a close correspondence was found between $T_2$ distributions of fixed [44] and *in-vivo* [45] rat SCs, but a different fixative was used in our study and different chemical preparations are known to affect exchange properties differently [73]. Therefore, this study should be followed by an *in-vivo* experiments to validate the phenomenon in such settings. In addition, because the main goal of this study was to explore the relationship $MET_2$ and MT for a broader understanding of the mechanisms involved, temporal resolution was not a limitation (the total acquisition time was ~30h per dataset). Reduction of the acquisition time (*e.g.* by reducing spectral resolution) would be required for an *in-vivo* preparation.

### 4.3. Framing MT and $MET_2$ in MR-based WM characterization

Looking beyond $MET_2$ and MT, many different MR approaches have been proposed to characterize WM. For example, diffusion-based MR techniques such as q-space imaging [5] or oscillating gradients spin echo [6] were shown to be valuable techniques to map WM characteristics as axon diameters, although requiring powerful gradients. Alternatively, axonal mapping $T_2^*$-based techniques have also been proposed to overcome this limitation [57] although only parallel fibers were considered. These techniques provide important WM information but don't address myelin content quantification. Multi-exponential $T_2^*$, similarly to $MET_2$, showed already promises to map myelin [74]; but such technique, not only is reported to be challenging at UHF [46] but also introduces a new unknown (additional dephasing in the transverse plan), and thus it may be more difficult to interpret results. Quantification of $T_1$ was also proposed, but its accurate measurement is challenging [75] and off-resonance irradiation was found to generate unexpected extra signals on $T_1$ measurements [76]. Multidimensionality has also been applied previously in a diffusion-relaxation correlation spectroscopy imaging study [77] that demonstrated the power of hybrid studies, in this case, for resolving tissue compartmentalization. Adding a third dimension (diffusion) may be an interesting follow-up of the present study. However, before further explorations, both histology and careful theoretical treatment are required to elucidate the observed mechanism.

## 5. Conclusion

In this study, the effects of irradiation at different offset frequencies on the MET$_2$ spectra were explored. MET$_2$ parameters were found to shift with off-resonance saturation, showing unique tract-specific signatures across the irradiation frequency. Offset-saturation-induced (osi-) MET$_2$ parametric maps show contrasts between microstructurally-distinct rat spinal cord tracts. Furthermore, the potential of exploiting the osi-MET$_2$ shift phenomenon to increase white matter contrast was demonstrated. Exchange of water or intra-axonal macromolecules have been suggested as hypothetical explanations for the source of the contrast, and further experiments are required to establish or refute either interpretation. Nevertheless. MT-MET$_2$ and the ensuing osi-shifts can add a new dimension for WM characterization that will greatly assist in investigating the source of the phenomena presented here.

## Acknowledgments

This study was supported through funding from European Research Council (ERC) under the European Union's Horizon 2020 research and innovation programme (Starting Grant, agreement No. 679058). TSD acknowledges funding from the Fundação da Ciência e Tecnologia (FCT), Portugal (SFRH/BD/124637/2016). Both authors would like to thank Dr. Daniel Nunes for the insightful discussions and providing the spinal cords used in this study. The authors are also grateful to Prof. Mark D Does and Dr. Kevin Harkins from Vanderbilt University for the REMMI pulse sequence and its analysis tools, that were supported through grant number NIH EB019980.

# References

[1]  E.R. Kandel, J.H. Schwartz, T.M. Jessell, Principles of Neural Sciences, 4th Editio, Chicago: Appleton&Lange, 2000.

[2]  R. Caminiti, F. Carducci, C. Piervincenzi, A. Battaglia-Mayer, G. Confalone, F. Visco-Comandini, P. Pantano, G.M. Innocenti, Diameter, Length, Speed, and Conduction Delay of Callosal Axons in Macaque Monkeys and Humans: Comparing Data from Histology and Magnetic Resonance Imaging Diffusion Tractography, J. Neurosci. 33 (2013) 14501–14511. doi:10.1523/JNEUROSCI.0761-13.2013.

[3]  J.A. Perge, J.E. Niven, E. Mugnaini, V. Balasubramanian, P. Sterling, Why Do Axons Differ in Caliber?, J. Neurosci. 32 (2012) 626–638. doi:10.1523/JNEUROSCI.4254-11.2012.

[4]  G.M. Innocenti, A. Vercelli, R. Caminiti, The diameter of cortical axons depends both on the area of origin and target, Cereb. Cortex. 24 (2014) 2178–2188. doi:10.1093/cercor/bht070.

[5]  H.H. Ong, A.C. Wright, S.L. Wehrli, A. Souza, E.D. Schwartz, S.N. Hwang, F.W. Wehrli, Indirect measurement of regional axon diameter in excised mouse spinal cord with q-space imaging: Simulation and experimental studies, Neuroimage. 40 (2008) 1619–1632. doi:10.1016/j.neuroimage.2008.01.017.

[6]  J. Xu, H. Li, K.D. Harkins, X. Jiang, J. Xie, H. Kang, M.D. Does, J.C. Gore, Mapping mean axon diameter and axonal volume fraction by MRI using temporal diffusion spectroscopy, Neuroimage. 103 (2014) 10–19. doi:10.1016/j.neuroimage.2014.09.006.

[7]  N. Shemesh, G.A. Alvarez, L. Frydman, Size distribution imaging by non-uniform oscillating-gradient spin echo (NOGSE) MRI, PLoS One. 10 (2015) 1–19. doi:10.1371/journal.pone.0133201.

[8]  J. Veraart, E. Fieremans, D.S. Novikov, Universal power-law scaling of water diffusion in human brain defines what we see with MRI, Prepr. ArXiv1609.09145. (2016) 1–8. http://arxiv.org/abs/1609.09145.

[9]  C. Laule, E. Leung, D.K. Li, A.L. Traboulsee, D.W. Paty, A.L. MacKay, G.R. Moore, Myelin water imaging in multiple sclerosis: quantitative correlations with histopathology, Mult. Scler. J. 12 (2006) 747–753. doi:10.1177/1352458506070928.

[10] K. Schmierer, D.J. Tozer, F. Scaravilli, D.R. Altmann, G.J. Barker, P.S. Tofts, D.H. Miller, Quantitative magnetization transfer imaging in postmortem multiple sclerosis brain, J. Magn. Reson. Imaging. 26 (2007) 41–51. doi:10.1002/jmri.20984.

[11] C.R. McCreary, T.A. Bjarnason, V. Skihar, J.R. Mitchell, V.W. Yong, J.F. Dunn, Multiexponential T2 and magnetization transfer MRI of demyelination and remyelination in murine spinal cord, Neuroimage. 45 (2009) 1173–1182. doi:10.1016/j.neuroimage.2008.12.071.

[12] A. MacKay, C. Laule, I. Vavasour, T. Bjarnason, S. Kolind, B. M??dler, Insights into brain microstructure from the T2 distribution, Magn. Reson. Imaging. 24

(2006) 515–525. doi:10.1016/j.mri.2005.12.037.

[13] C. Laule, I.M. Vavasour, G.R.W. Moore, J. Oger, D.K.B. Li, D.W. Paty, A.L. MacKay, Water content and myelin water fraction in multiple sclerosis: A T2 relaxation study, J. Neurol. 251 (2004) 284–293. doi:10.1007/s00415-004-0306-6.

[14] R.I. Grossman, Magnetization transfer in multiple sclerosis, Ann Neurol. 36 Suppl (1994) S97-9. http://www.ncbi.nlm.nih.gov/pubmed/8017897.

[15] K. Schmierer, F. Scaravilli, D.R. Altmann, G.J. Barker, D.H. Miller, Magnetization transfer ratio and myelin in postmortem multiple sclerosis brain, Ann. Neurol. 56 (2004) 407–415. doi:10.1002/ana.20202.

[16] E. Van Obberghen, S. Mchinda, A. Le Troter, V.H. Prevost, P. Viout, M. Guye, G. Varma, D.C. Alsop, J.P. Ranjeva, J. Pelletier, O. Girard, G. Duhamel, Evaluation of the sensitivity of inhomogeneous magnetization transfer (ihMT) MRI for multiple sclerosis, Am. J. Neuroradiol. 39 (2018) 634–641. doi:10.3174/ajnr.A5563.

[17] N. Stikov, J.S.W. Campbell, T. Stroh, M. Lavelée, S. Frey, J. Novek, S. Nuara, M.-K. Ho, B.J. Bedell, R.F. Dougherty, I.R. Leppert, M. Boudreau, S. Narayanan, T. Duval, J. Cohen-Adad, P. Picard, A. Gasecka, D. Côté, G. Bruce Pike, In vivo histology of the myelin g-ratio with magnetic resonance imaging., Neuroimage. 118 (2015) 397–405. doi:10.1016/j.neuroimage.2015.05.023.

[18] A. Mackay, K. Whittall, J. Adler, D. Li, D. Paty, D. Graeb, In vivo visualization of myelin water in brain by MR.pdf, Magn Reson Med. (1994).

[19] R.S. Menon, P.S. Allen, Application of continuous relaxation time distributions to the fitting of data from model systems and excised tissue., Magn. Reson. Med. 20 (1991) 214–227. doi:10.1002/mrm.1910200205.

[20] K.P. Whittall, A.L. Mackay, D.A. Graeb, R.A. Nugent, D.K.B. Li, D.W. Paty, In vivo measurement of T2 distributions and water contents in normal human brain, Magn. Reson. Med. 37 (1997) 34–43. doi:10.1002/mrm.1910370107.

[21] C. Laule, I.M. Vavasour, S.H. Kolind, D.K.B. Li, T.L. Traboulsee, G.R.W. Moore, A.L. MacKay, Magnetic Resonance Imaging of Myelin, Neurotherapeutics. 4 (2007) 460–484. doi:10.1016/j.nurt.2007.05.004.

[22] E. Alonso-Ortiz, I.R. Levesque, G.B. Pike, MRI-based myelin water imaging: A technical review, Magn. Reson. Med. 73 (2015) 70–81. doi:10.1002/mrm.25198.

[23] R.D. Dortch, K.D. Harkins, M.R. Juttukonda, J.C. Gore, M.D. Does, Characterizing inter-compartmental water exchange in myelinated tissue using relaxation exchange spectroscopy, Magn. Reson. Med. 70 (2013) 1450–1459. doi:10.1002/mrm.24571.

[24] C.L. Lawson, R.J. Hanson, Solving least squares problems, Prentice-Hall, 1974.

[25] S.W. Provencher, A constrained regularization method for inverting data represented by linear algebraic or integral equations, Comput. Phys. Commun. 27 (1982) 213–227. doi:10.1016/0010-4655(82)90173-4.

[26] G.. Borgia, R.J.. Brown, P. Fantazzini, Uniform-Penalty Inversion of Multiexponential Decay Data, J. Magn. Reson. 147 (2000) 273–285.

doi:10.1006/jmre.2000.2197.

[27] M.D. Does, R.E. Snyder, Multiexponential T2 relaxation in degenerating peripheral nerve, Magn. Reson. Med. 35 (1996) 207–213. doi:10.1002/mrm.1910350212.

[28] P. Kozlowski, J. Liu, A.C. Yung, W. Tetzlaff, High-resolution myelin water measurements in rat spinal cord, Magn. Reson. Med. 59 (2008) 796–802. doi:10.1002/mrm.21527.

[29] W. Oakden, J.M. Kwiecien, M.A. O'Reilly, W. Dabrowski, C. Whyne, J. Finkelstein, K. Hynynen, G.J. Stanisz, Quantitative MRI in a non-surgical model of cervical spinal cord injury, NMR Biomed. 28 (2015) 925–936. doi:10.1002/nbm.3326.

[30] H.S.-M. Chen, N. Holmes, J. Liu, W. Tetzlaff, P. Kozlowski, Validating myelin water imaging with transmission electron microscopy in a rat spinal cord injury model, Neuroimage. 153 (2017) 122–130. doi:10.1016/j.neuroimage.2017.03.065.

[31] S.D. Wolff, R.S. Balaban, Magnetization transfer contrast (MTC) and tissue water proton relaxation in vivo, Magn Reson Med. 10 (1989) 135–144. doi:10.1002/mrm.1910100113.

[32] R.M. Henkelman, X. Huang, Q.-S.S. Xiang, G.J. Stanisz, S.D. Swanson, M.J. Bronskill, Quantitative interpretation of magnetization transfer, Magn Reson Med. 29 (1993) 759–766. doi:10.1002/mrm.1910290607.

[33] D.F. Gochberg, J.C. Gore, Quantitative magnetization transfer imaging via selective inversion recovery with short repetition times, Magn. Reson. Med. 57 (2007) 437–441. doi:10.1002/mrm.21143.

[34] S.D. Wolff, R.S. Balaban, Magnetization transfer imaging: practical aspects and clinical applications., Radiology. 192 (1994) 593–599. doi:10.1148/radiology.192.3.8058919.

[35] C. Morrison, R. Mark Henkelman, A Model for Magnetization Transfer in Tissues, Magn. Reson. Med. 33 (1995) 475–482. doi:10.1002/mrm.1910330404.

[36] J. Lee, R.R. Regatte, A. Jerschow, Isolating chemical exchange saturation transfer contrast from magnetization transfer asymmetry under two-frequency rf irradiation, J. Magn. Reson. 215 (2012) 56–63. doi:10.1016/j.jmr.2011.12.012.

[37] G. Liu, X. Song, K.W.Y. Chan, M.T. Mcmahon, Nuts and bolts of chemical exchange saturation transfer MRI, (2013) 810–828. doi:10.1002/nbm.2899.

[38] J. Rademacher, V. Engelbrecht, U. Bu, H. Freund, K. Zilles, Measuring in Vivo Myelination of Human White Matter Fiber Tracts with Magnetization Transfer MR, 406 (1999) 393–406.

[39] J.P. Mottershead, K. Schmierer, M. Clemence, J.S. Thornton, F. Scaravilli, G.J. Barker, P.S. Tofts, J. Newcombe, M.L. Cuzner, R.J. Ordidge, W.I. Mcdonald, D.H. Miller, High field MRI correlates of myelin content and axonal density in multiple sclerosis A post-mortem study of the spinal cord, J Neurol. (2003) 1293–1301. doi:10.1007/s00415-003-0192-3.

[40] W. Zaaraoui, M. Deloire, M. Merle, C. Girard, G. Raffard, M. Biran, M. Inglese, K.

Petry, O. Gonen, B. Brochet, J.-M. Franconi, D. Vicent, Monitoring demyination and remyelization transfer imaging in the mouse brain at 9.4 T, MAGMA. 21 (2008) 357–362. doi:10.1007/s10334-008-0141-3.Monitoring.

[41] M.S.A. Deloire-Grassin, B. Brochet, B. Quesson, C. Delalande, V. Dousset, P. Canioni, K.G. Petry, In vivo evaluation of remyelination in rat brain by magnetization transfer imaging, J. Neurol. Sci. 178 (2000) 10–16. doi:10.1016/S0022-510X(00)00331-2.

[42] I.M. Vavasour, K.P. Whittall, A.L. MacKay, D.K.B. Li, G. Vorobeychik, D.W. Paty, A comparison between magnetization transfer ratios and myelin water percentages in normals and multiple sclerosis patients, Magn. Reson. Med. 40 (1998) 763–768. doi:10.1002/mrm.1910400518.

[43] D.J. Tozer, G.R. Davies, D.R. Altmann, D.H. Miller, P.S. Tofts, Correlation of apparent myelin measures obtained in multiple sclerosis patients and controls from magnetization transfer and multicompartmental T 2 analysis, Magn. Reson. Med. 53 (2005) 1415–1422. doi:10.1002/mrm.20479.

[44] A.N. Dula, D.F. Gochberg, H.L. Valentine, W.M. Valentine, M.D. Does, Multiexponential T2, magnetization transfer, and Quantitative histology in white matter tracts of rat spinal cord, Magn. Reson. Med. 63 (2010) 902–909. doi:10.1002/mrm.22267.

[45] K.D. Harkins, A.N. Dula, M.D. Does, Effect of intercompartmental water exchange on the apparent myelin water fraction in multiexponential T 2 measurements of rat spinal cord, Magn. Reson. Med. 67 (2012) 793–800. doi:10.1002/mrm.23053.

[46] K.L. West, N.D. Kelm, R.P. Carson, D.F. Gochberg, K.C. Ess, M.D. Does, Myelin volume fraction imaging with MRI, Neuroimage. (2016) 0–1. doi:10.1016/j.neuroimage.2016.12.067.

[47] A. Bax, L. Lerner, Two-Dimensional Nuclear Magnetic Resonance Spectroscopy, Science (80-. ). 232 (1986) 960–967.

[48] R. Harrison, M.J. Bronskill, R.M. Henkelman, Magnetization Transfer and T2 Relaxation Components in Tissue, Magn Reson Med. (1995) 490–496.

[49] G.J. Stanisz, A. Kecojevic, M.J. Bronskill, R.M. Henkelman, Characterizing White Matter With Magnetization Transfer and T2, Magn. Reson. Med. 1136 (1999) 1128–1136.

[50] I.M. Vavasour, K.P. Whittall, D.K. Li,  a L. MacKay, Different magnetization transfer effects exhibited by the short and long T(2) components in human brain., Magn. Reson. Med. 44 (2000) 860–6. doi:10.1002/1522-2594(200012)44:6<860::AID-MRM6>3.0.CO;2-C.

[51] R. Pohmann, G. Shajan, D.Z. Balla, Contrast at High Field : Relaxation Times, Magnetization Transfer and Phase in the Rat Brain at 16.4 T, Magn Reson Med. 1581 (2011) 1572–1581. doi:10.1002/mrm.22949.

[52] C. Eichner, S.F. Cauley, J. Cohen-Adad, H.E. Möller, R. Turner, K. Setsompop, L.L. Wald, Real diffusion-weighted MRI enabling true signal averaging and increased diffusion contrast, Neuroimage. 122 (2015) 373–384. doi:10.1016/j.neuroimage.2015.07.074.

[53] J. Veraart, D.S. Novikov, D. Christiaens, B. Ades-aron, J. Sijbers, E. Fieremans, Denoising of diffusion MRI using random matrix theory, Neuroimage. 142 (2016) 394–406. doi:10.1016/j.neuroimage.2016.08.016.

[54] E. Kellner, B. Dhital, V.G. Kiselev, M. Reisert, Gibbs-ringing artifact removal based on local subvoxel-shifts, Magn. Reson. Med. 76 (2016) 1574–1581. doi:10.1002/mrm.26054.

[55] G. Paxinos, C. Watson, The Rat Brain in Stereotaxic Coordinates, 5th Editio, Academic Press, 2004. doi:10.1017/CBO9781107415324.004.

[56] A. Nogradi, G. Vrbova, G. Anatomy, Transplantation of Neural Tissue into the Spinal Cord ;Anatomy and Physiology of the Spinal Cord, in: Transplant. Neural Tissue into Spinal Cord, 2006: pp. 1–23.

[57] D. Nunes, T.L. Cruz, S.N. Jespersen, N. Shemesh, Mapping axonal density and average diameter using non-monotonic time-dependent gradient-echo MRI, J. Magn. Reson. 277 (2017) 117–130. doi:10.1016/j.jmr.2017.02.017.

[58] E.D. Schwartz, E.T. Cooper, C.-L. Chin, S. Wehrli, A. Tessler, D.B. Hackney, Ex vivo evaluation of ADC values within spinal cord white matter tracts, AJNR. Am. J. Neuroradiol. 26 (2005) 390–397.

[59] R.M. Henkelman, G.J. Stanisz, S.J. Graham, Magnetization transfer in MRI : a review, (2001) 57–64. doi:10.1002/nbm.683.

[60] J. Hua, C.K. Jones, J. Blakeley, S.A. Smith, P.C.M. Van Zijl, J. Zhou, Quantitative Description of the Asymmetry in Magnetization Transfer Effects Around the Water Resonance in the Human Brain, 793 (2007) 786–793. doi:10.1002/mrm.21387.

[61] J. Kim, J. Choi, J. Cho, C. Lee, D. Jeon, S. Park, Preliminary Observations on Sensitivity and Specificity of Magnetization Transfer Asymmetry for Imaging Myelin of Rat Brain at High Field, 2015 (2015) 12–17. doi:10.1155/2015/565391.

[62] K.P. Whittall, A.L. MacKay, Quantitative interpretation of NMR relaxation data, J. Magn. Reson. 84 (1989) 134–152. doi:10.1016/0022-2364(89)90011-5.

[63] T. Prasloski, B. Mädler, Q.S. Xiang, A. MacKay, C. Jones, Applications of stimulated echo correction to multicomponent T2analysis, Magn. Reson. Med. 67 (2012) 1803–1814. doi:10.1002/mrm.23157.

[64] S.D. Swanson, P. Y., MT is Symmetric but Shifted with Respect to Water, Proc. Intl. Soc. Mag. Reson. Med. 35 (2003) 2003.

[65] P.C.M. Van Zijl, N.N. Yadav, Chemical Exchange Saturation Transfer ( CEST ): What is in a Name and What Isn ' t ?, 948 (2011) 927–948. doi:10.1002/mrm.22761.

[66] X. Zhang, F. Wang, T. Jin, J. Xu, J. Xie, D.F. Gochberg, J.C. Gore, Z. Zu, MR Imaging of a Novel NOE-Mediated Magnetization Transfer with Water in Rat Brain at 9 . 4 T, 00 (2016) 1–10. doi:10.1002/mrm.26396.

[67] M.D. Does, C. Beaulieu, P.S. Allen, R.E. Snyder, Multi-Component T1 Relaxation and Magnetisation Transfer in Peripheral Nerve, Magn. Reson. Imaging. 16 (1998) 1033–1041.

[68] R. Gil, D. Khabipova, M. Zwiers, T. Hilbert, T. Kober, J.P. Marques, An in vivo

study of the orientation-dependent and independent components of transverse relaxation rates in white matter, NMR Biomed. 29 (2016) 1780–1790. doi:10.1002/nbm.3616.

[69]  R. Stollberger, P. Wach, Imaging of the active B1 field in vivo, Magn. Reson. Med. 35 (1996) 246–251. doi:10.1002/mrm.1910380225.

[70]  W. Oakden, G.J. Stanisz, Effects of diffusion on high-resolution quantitative T2 MRI, NMR Biomed. 27 (2014) 672–680. doi:10.1002/nbm.3104.

[71]  G.A. Álvarez, N. Shemesh, L. Frydman, Internal gradient distributions: A susceptibility-derived tensor delivering morphologies by magnetic resonance, Sci. Rep. 7 (2017) 3311. doi:10.1038/s41598-017-03277-9.

[72]  H.E. D'Arceuil, S. Westmoreland, A.J. de Crespigny, An approach to high resolution diffusion tensor imaging in fixed primate brain, Neuroimage. 35 (2007) 553–565. doi:10.1016/j.neuroimage.2006.12.028.

[73]  T.M. Shepherd, P.E. Thelwall, G.J. Stanisz, S.J. Blackband, Aldehyde fixative solutions alter the water relaxation and diffusion properties of nervous tissue, Magn. Reson. Med. 62 (2009) 26–34. doi:10.1002/mrm.21977.

[74]  P. Sati, P. van Gelderen, A.C. Silva, D.S. Reich, H. Merkle, J.A. De Zwart, J.H. Duyn, Micro-compartment specific T2* relaxation in the brain, Neuroimage. 77 (2013) 268–278. doi:10.1016/j.neuroimage.2013.03.005.

[75]  P. van Gelderen, X. Jiang, J.H. Duyn, Rapid measurement of brain macromolecular proton fraction with transient saturation transfer MRI, Magn. Reson. Med. 77 (2017) 2174–2185. doi:10.1002/mrm.26304.

[76]  C. Duan, C. Ryan, S. Utsuzawa, Y.Q. Song, M.D. Hürlimann, Effect of off-resonance on T1 saturation recovery measurement in inhomogeneous fields, J. Magn. Reson. 281 (2017) 31–43. doi:10.1016/j.jmr.2017.05.003.

[77]  D. Kim, E.K. Doyle, J.L. Wisnowski, J.H. Kim, J.P. Haldar, Diffusion-relaxation correlation spectroscopic imaging: A multidimensional approach for probing microstructure, Magn. Reson. Med. 00 (2017). doi:10.1002/mrm.26629.

# Figures

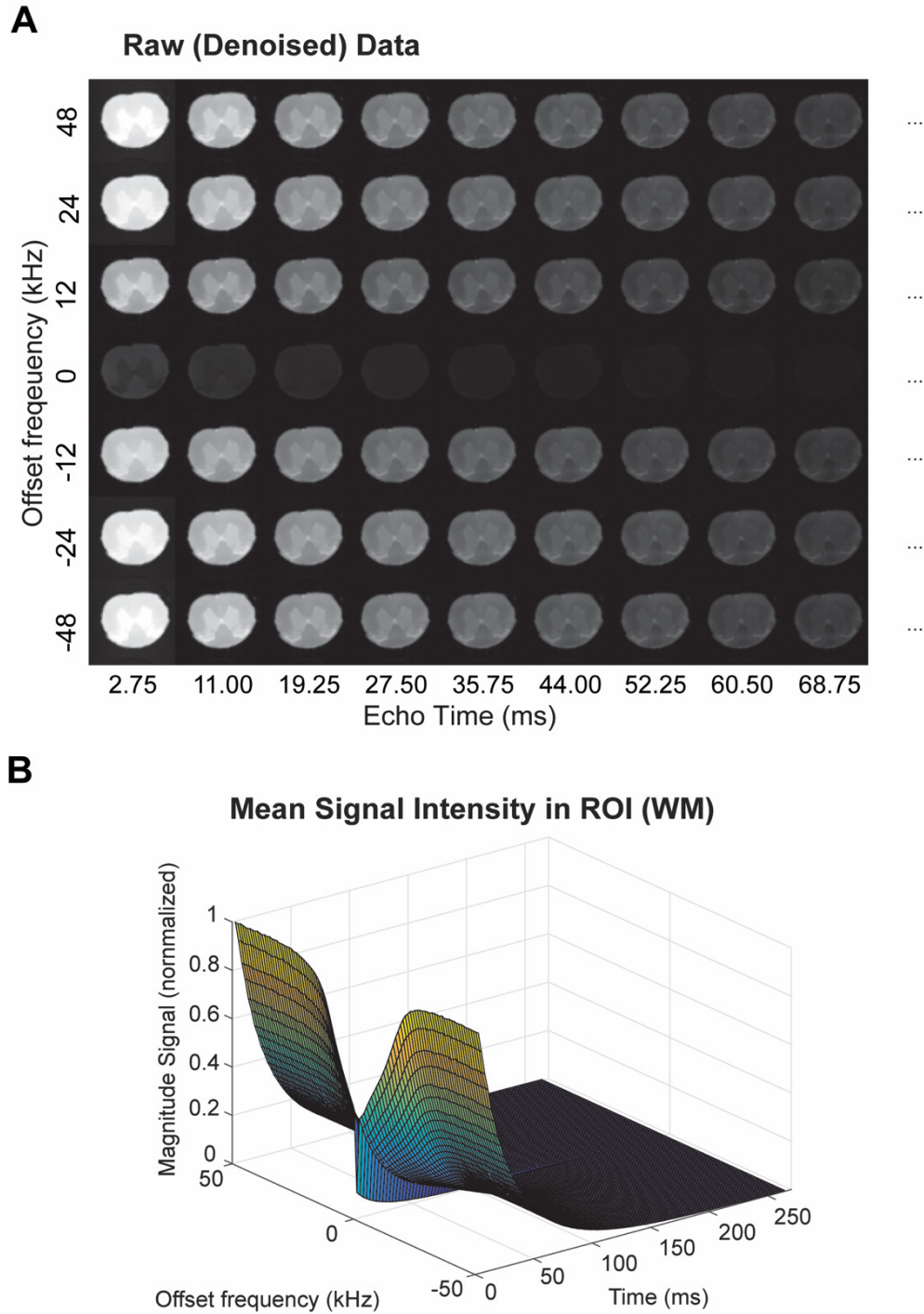

**Figure 1 - MT-MSME data.** A) Overall appearance of MT-MSE datasets (denoised data) from one representative spinal cord (SC #1). Denoised images at different offset frequencies and echo times, representative of a full dataset. **B)** Representation of pre-processed signal (as observed in B) for one large ROI in white matter. Clear MT

modulation can be depicted along one dimension (offset frequency) while $T_2$ decay is observed along the other (echo time).

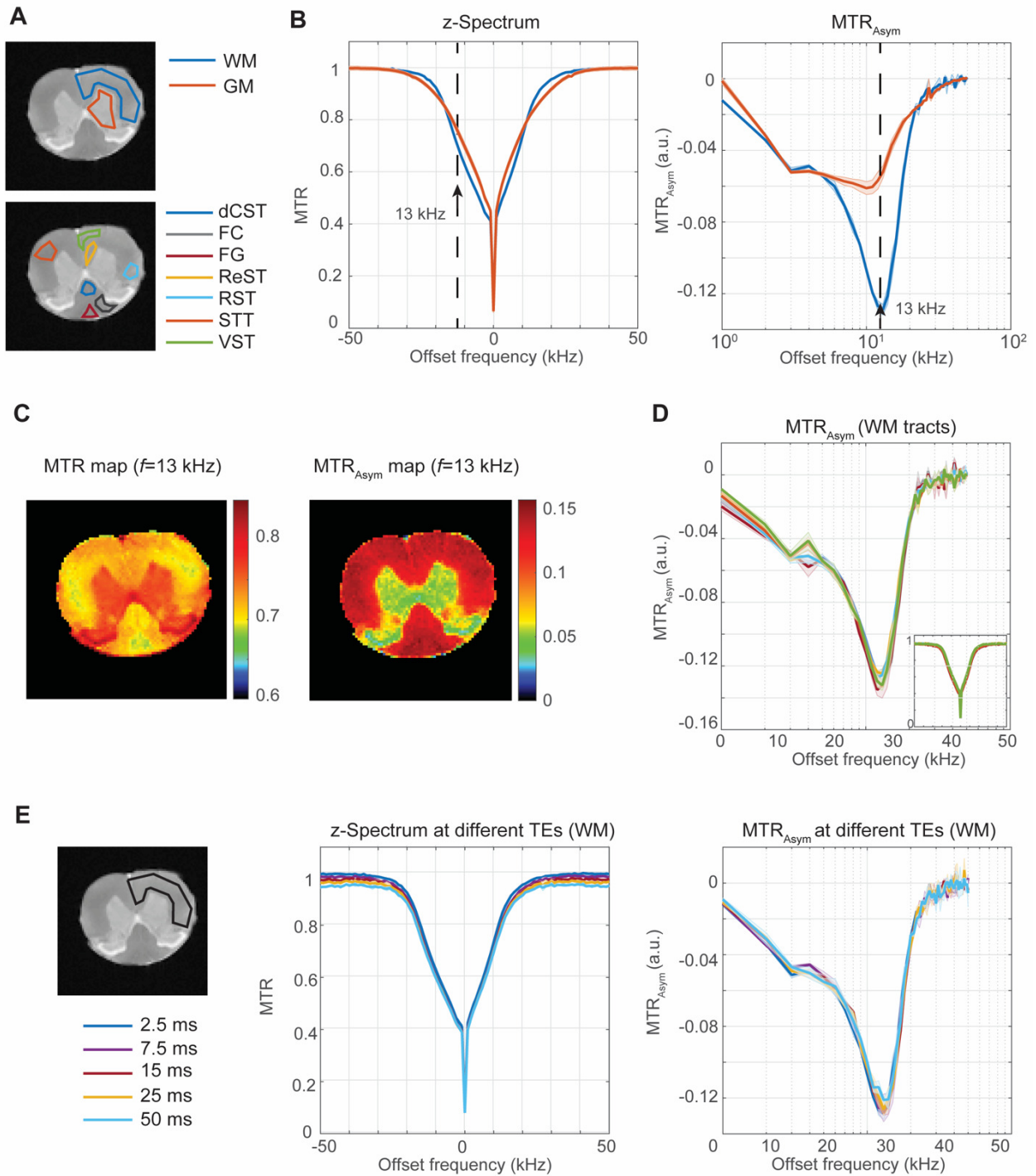

**Figure 2 - MT analysis (MTR and MTR$_{asym}$).** Plots display the Mean±SEM among SCs. The x-axis of MTR$_{asym}$ plots are shown in logarithmic scale. **A)** Colour legend of the ROIs analysed. Top: WM vs GM; bottom: seven WM tracts. **B)** z-spectrum and MTR$_{asym}$ plot

in WM vs GM (TE=2.75 ms). **C)** MTR and MTR$_{asym}$ maps of a representative SC (SC #1) at frequencies f=-13 kHz and f=13kHz, respectively (TE=2.75 ms). These frequencies are highlighted with arrow in z-spectrum and MTR$_{asym}$ plot of panel B). Substantial WM/GM contrast is observed whereas most WM tracts cannot be distinguished. **D)** MTR$_{asym}$ plot and z-spectrum (inset) in seven different WM tracts (TE=2.75 ms). Only minor changes are shown. **E)** z-spectrum and MTR$_{asym}$ plot at different echoes in WM. Though the position of asymptote in the z-spectra changes with TE (due to SNR loss), changes in the MT modulation are not perceived.

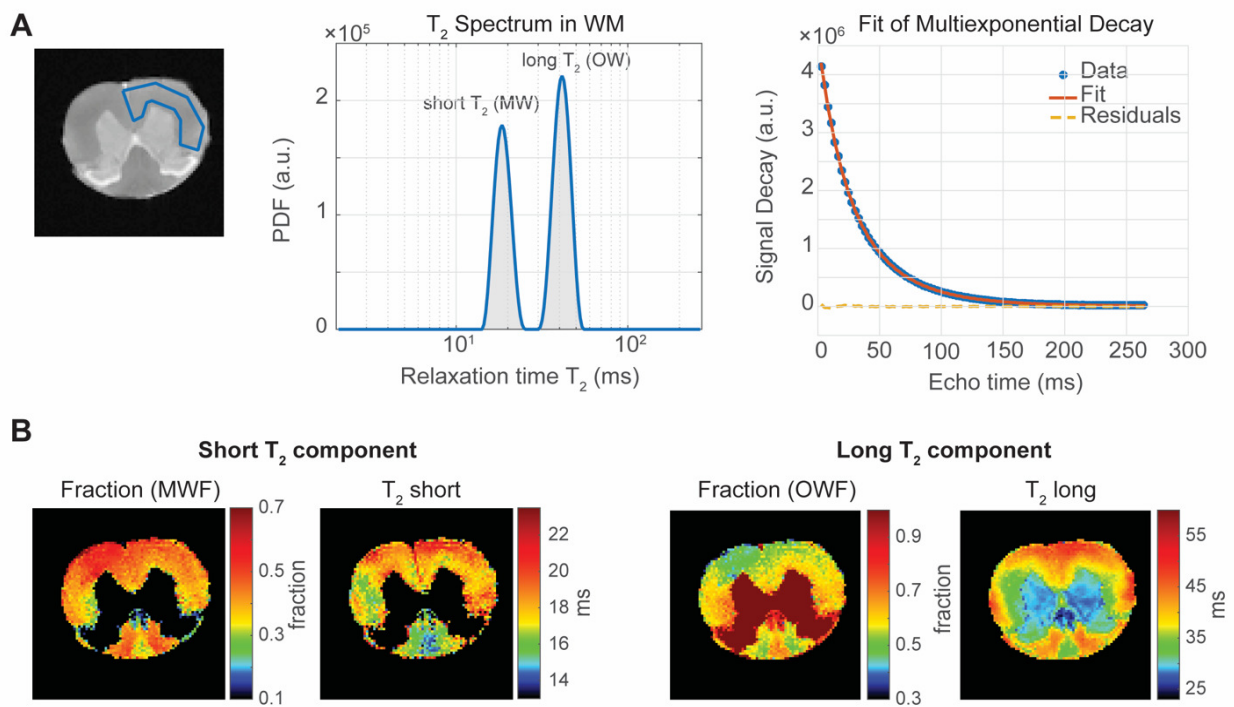

**Figure 3– Analysis of MET$_2$ decays by Laplace inversion of one representative SC (SC #1). A)** Analysis in a large ROI in WM. Left: Position of WM ROI. Centre: T$_2$s distribution. The two peaks are highlighted. Right: Comparison of acquired data with the inversion of products from the inverse Laplace analysis. The residuals of the fit are also plotted. The simulated decay fits well the data. **B)** MET$_2$ parametric maps obtain by voxelwise Laplace inversion. Most maps show spatially distinct contrast within the WM.

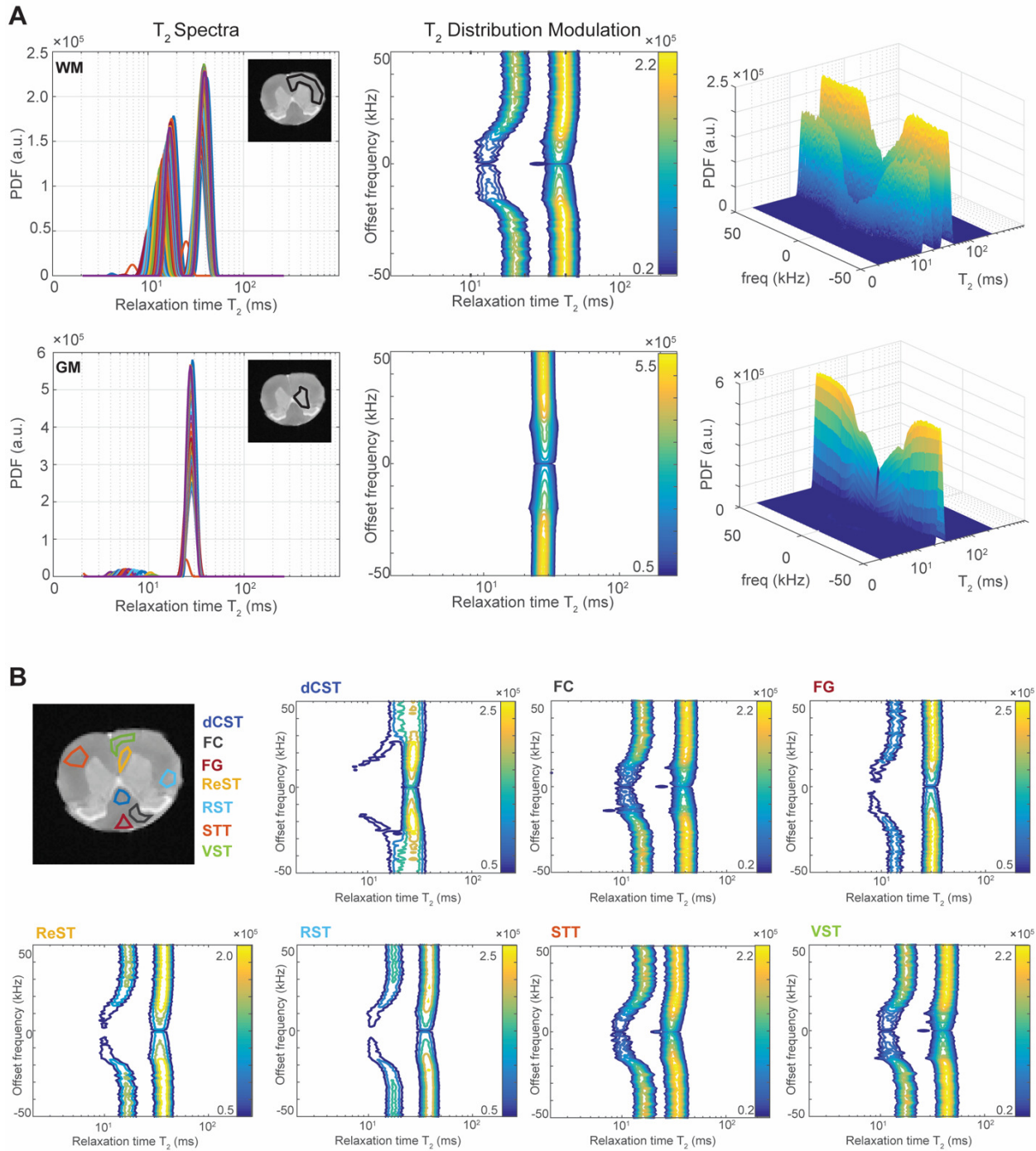

**Figure 4 – T$_2$ distributions over offset frequency in WM of one representative SC (SC #2). A)** T$_2$ distributions in WM (top) vs GM (bottom) presented in different views. Left: Different colours represent distributions at different offset frequencies. Centre: Contour map of T$_2$s against offset frequency. Right: Same modulation in a 3-D view. The two WM-peaks are shifting with offset irradiation. Such phenomenon is less evident in the single

GM peak. **B)** Contour maps of $T_2$ vs offset frequency in seven different WM tract ROIs. Both $T_2$ distributions and its modulations with offset frequency are tract-specific.

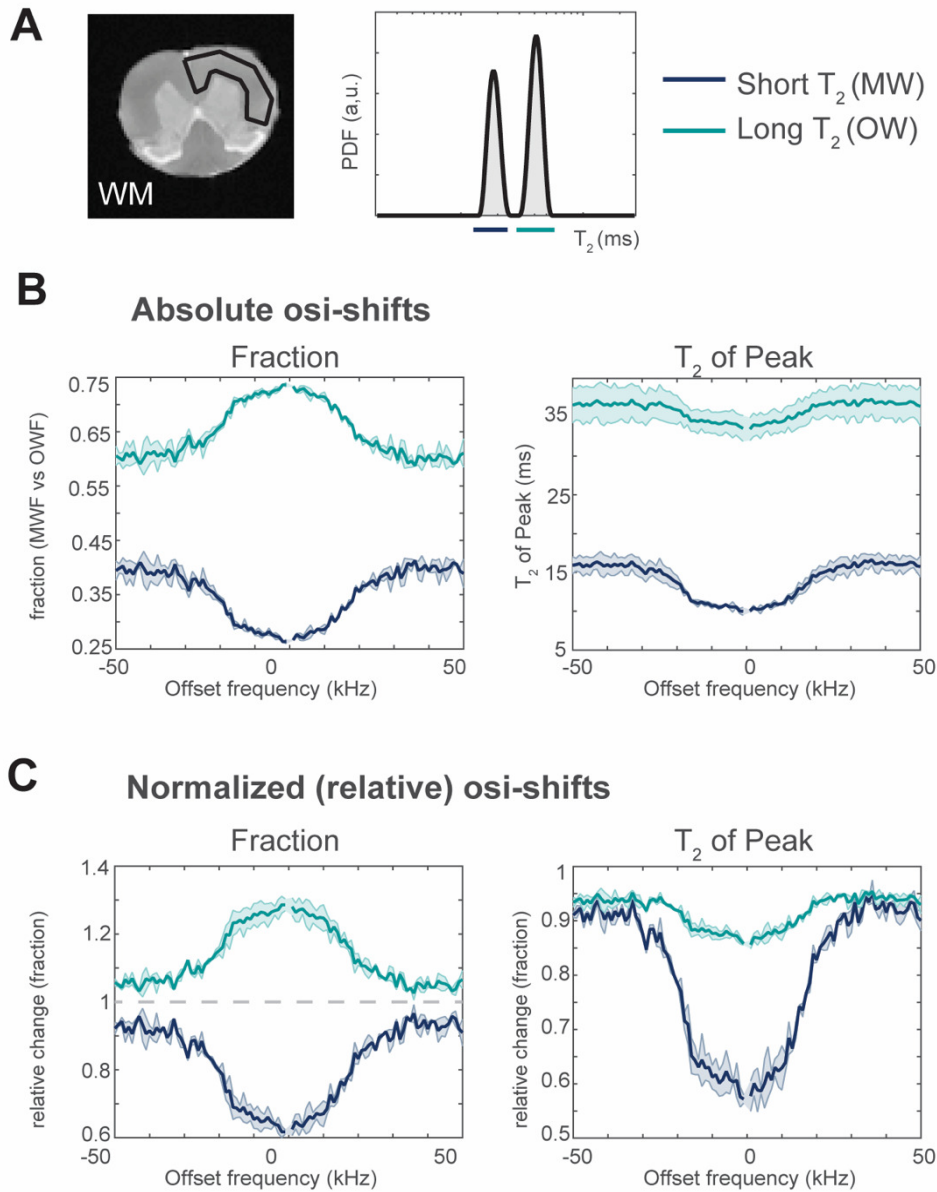

**Figure 5 - Offset-saturation-induced (osi-) shifts of MET$_2$ parameters from short and long T$_2$ WM components** (Mean±SEM among SCs). **A)** Illustration of WM ROI position and parameters extracted from one representative distribution. Short and long T$_2$ components are colour-coded. **B)** The fraction and T$_2$ of peak of short and long T$_2$ components over offset frequency. **C)** Relative change of plots in B) (normalized to the reference). Osi-shifts are observed in all parameters of both components, though the short $T_2$ component displays larger amplitude of modulation.

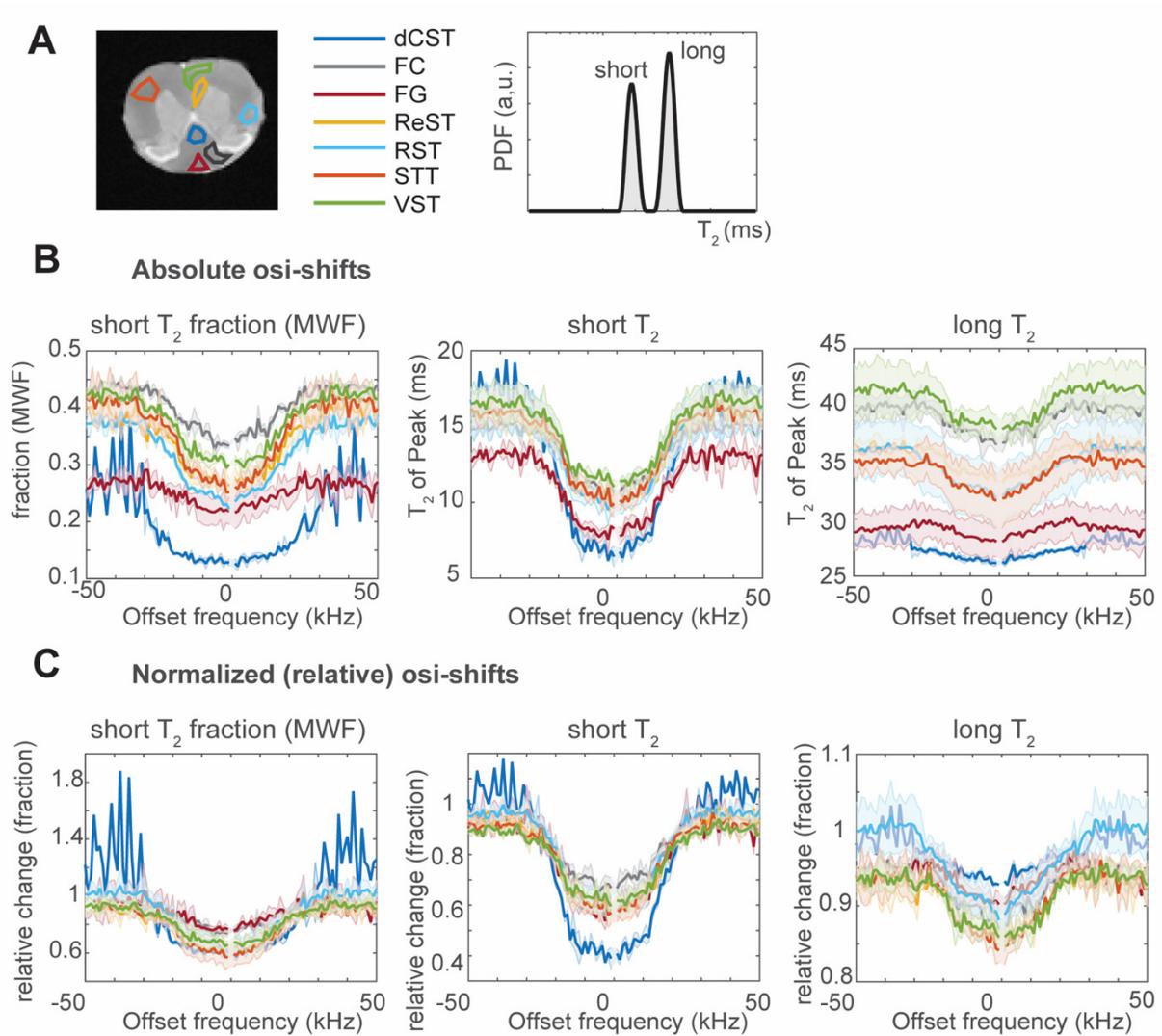

**Figure 6 – Offset-saturation-induced (osi-) shifts of MET$_2$ parameters of seven different WM tracts** (Mean±SEM among SCs). **A)** Illustration of ROIs position and parameters extracted from one representative distribution. The different WM tracts are colour-coded. The fractions and $T_2$s short and long inl **B** absolute and **C)** relative scales. Osi-shift of short and long $T_2$s are tracts-specific.

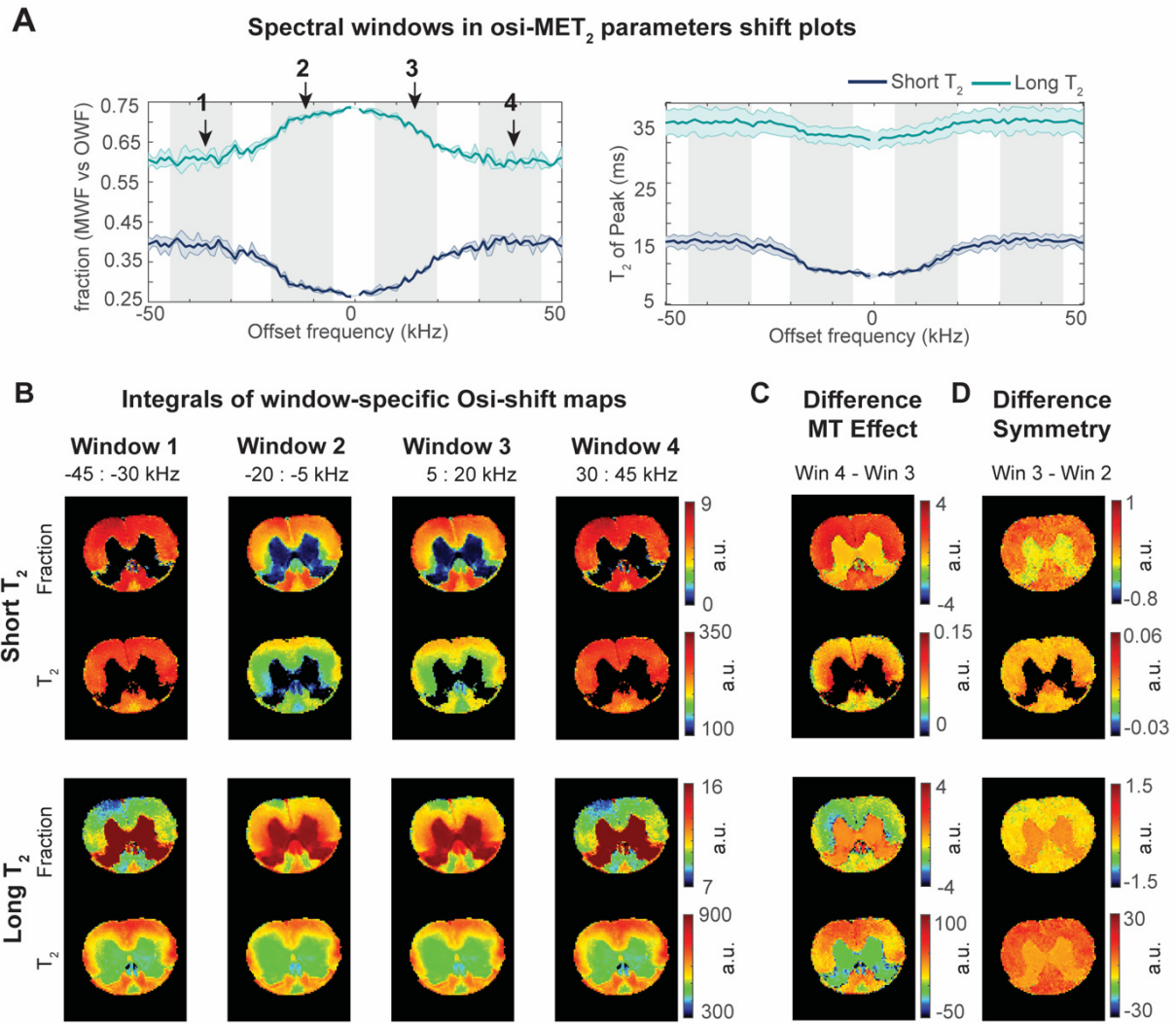

**Figure 7 - Voxelwise integration of region-specific osi-MET$_2$ parameters shifts of one representative SC** (SC #1) **A)** The range of frequencies or "spectral windows" integrated are illustrated as shades on the osi-shift plots of the same SC (WM ROI). The windows' numbering is marked with arrows in the fraction osi-shift plot. **B)** Maps of integrals of window-specific osi-shifts. The osi-shifts integrals are spatially-distinct. **C)** Difference map between a low MT-saturated window (Window 3) and a highly MT-saturated window (Window 4). Difference map displays distinct contrast from single region integral of osi-shift maps, which can be exploited to enhance contrast between WM tracts. **D)** Difference map between two symmetrical spectral windows (Window 3-Window 2). Both short and long T$_2$ parametric maps are homogeneous.

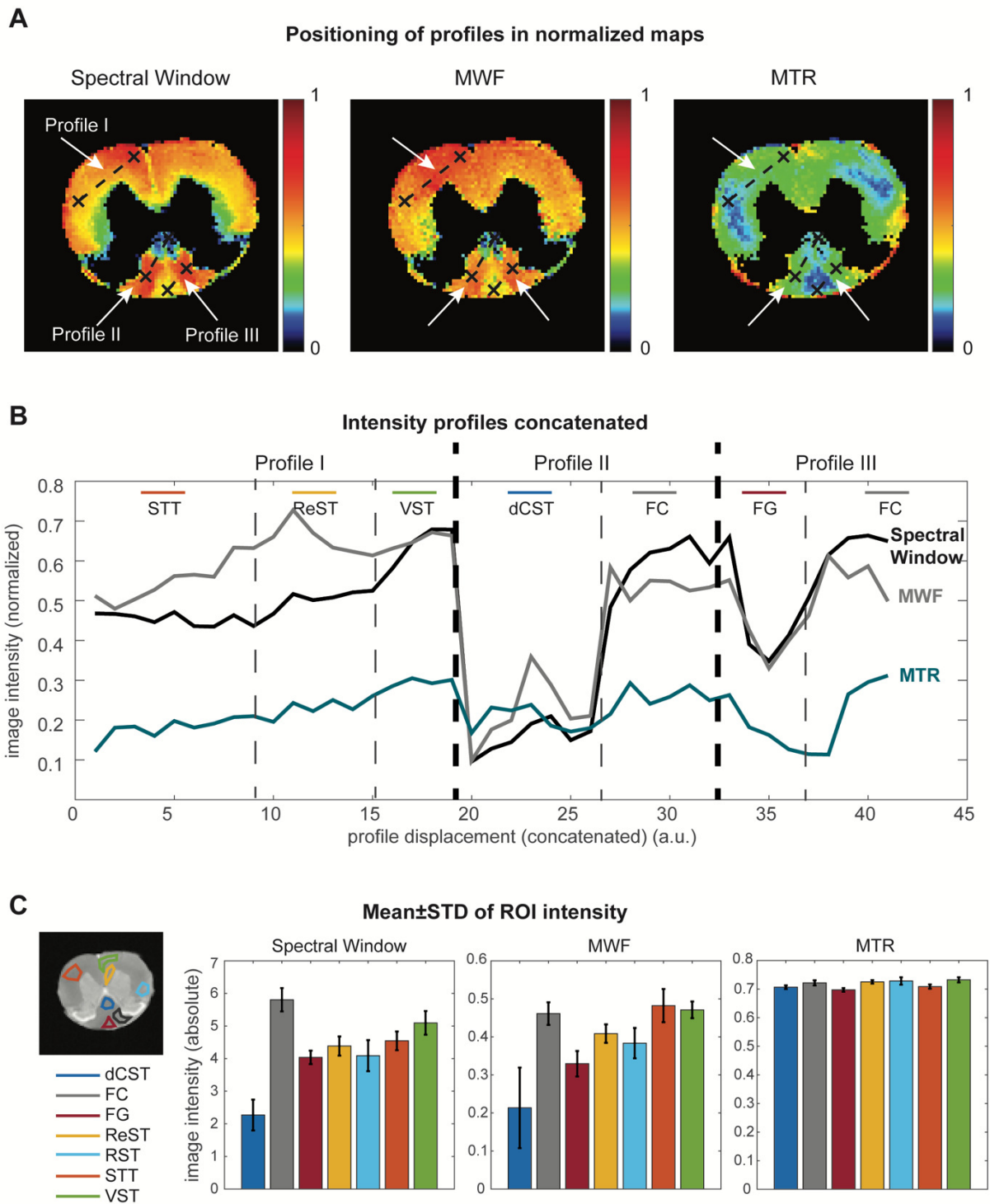

**Figure 8 – Enhancement of contrast provided by the Spectral Window maps. A)** Three profile lines were positioned along different WM tracts in the normalized Spectral Window, MWF and MTR maps. The profiles are marked with black dashed lines. White arrows are placed to help deciphering the profile lines. **B)** The normalized signal along the profiles for the three different techniques (Spectral window, MWF and MTR) are shown. Dashed lines indicate the separation between different WM tracts. **C)** Mean±STD of ROI intensity in absolute values are shown for each WM tract ROI and each technique (Spectral Window, MWF and MTR). Spectral Window profiles appear to be able to better distinguish different ROIs, appearing as a potential method to enhance contrast between WM tracts.

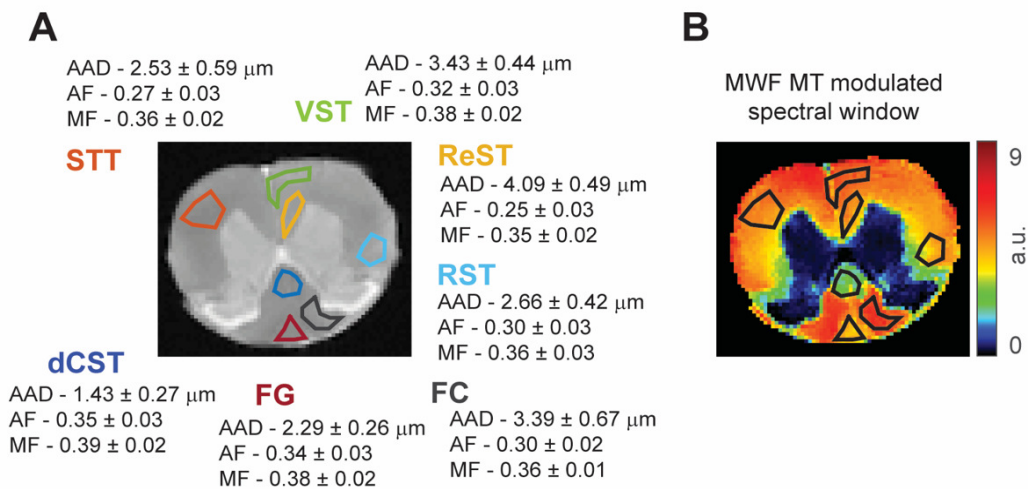

**Figure 9** –**Comparison of tract-specific histological features with contrast from osi-shift difference map of short T$_2$ component fraction (MWF). A)** WM tract ROIs position and corresponding histological features from a study conducted recently in the laboratory (Nunes et al., 2017). The spinal cords in this study were treated in a similar way. The features measured were: AAD – average axonal diameter (in µm); AF – axonal fraction; and MF – myelin fraction. **B)** MWF difference map from spectral window 4 (weak MT modulation) and window 3 (strong MT modulation) – see Figure 7 for clarification.

# Supplementary Figures

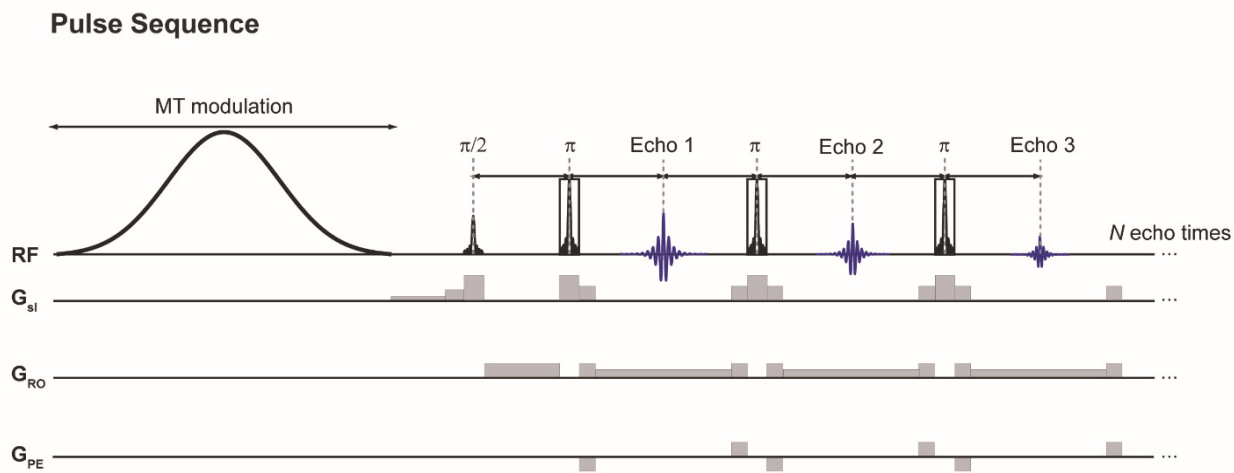

**Supplementary Figure 1- MT-MSME pulse sequence.** Pulse-sequence employed to generate two-dimensional MT-MET$_2$ datasets. A multi spin echo sequence was preceded by a gaussian-shaped 2 sec length RF pulse. MT offset frequencies were varied between {-50,50 kHz} in 1 kHz increments, plus one reference acquisition (without MT saturation).

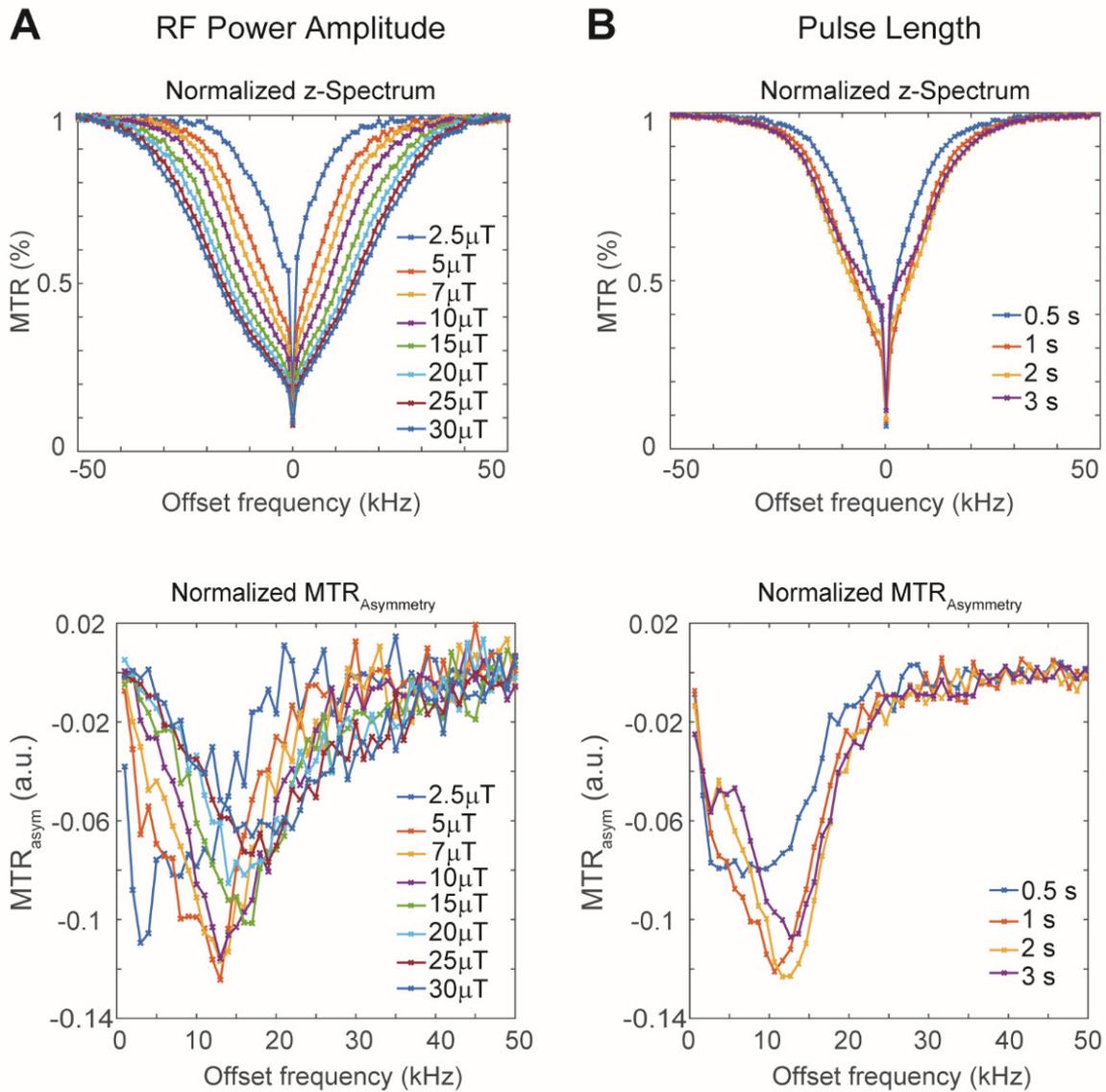

**Supplementary Figure 2 – Optimization of MT module parameters**. One additional spinal cord was used for this preliminary study. Single-shot EPIs with TR/TE=1050/26ms and res=(63x50x800)µm$^3$ were preceded by gaussian-shaped MT modulations with varying RF power amplitudes and lengths (offset frequencies=-50:1:50kHz). **A)** The normalized z-spectrum (top) and MTR$_{asym}$ plot (bottom) of MT modules with RF power amplitudes of 2.5, 5.0, 10.0, 15.0, 20.0, 25.0, 30.0, 35.0 µT were compared (fixed pulse length of 2 s). **B)** The previous experiment was repeated with fixed RF power amplitude =5 µT and extended TR = 3050 ms to allow the comparison of pulse lengths of 0.5, 1, 2.0, 3.0 s. If the macromolecular pool was the main culprit for centre frequency shift, then the highest sensitivity to myelin would be expected for the parameters with maximal

MTR$_{asym}$ effect. Such condition was observed with RF power of 5µT and pulse length of 2s. Similar parameters were used by others at 16.4 T (Pohmann et al., 2011).

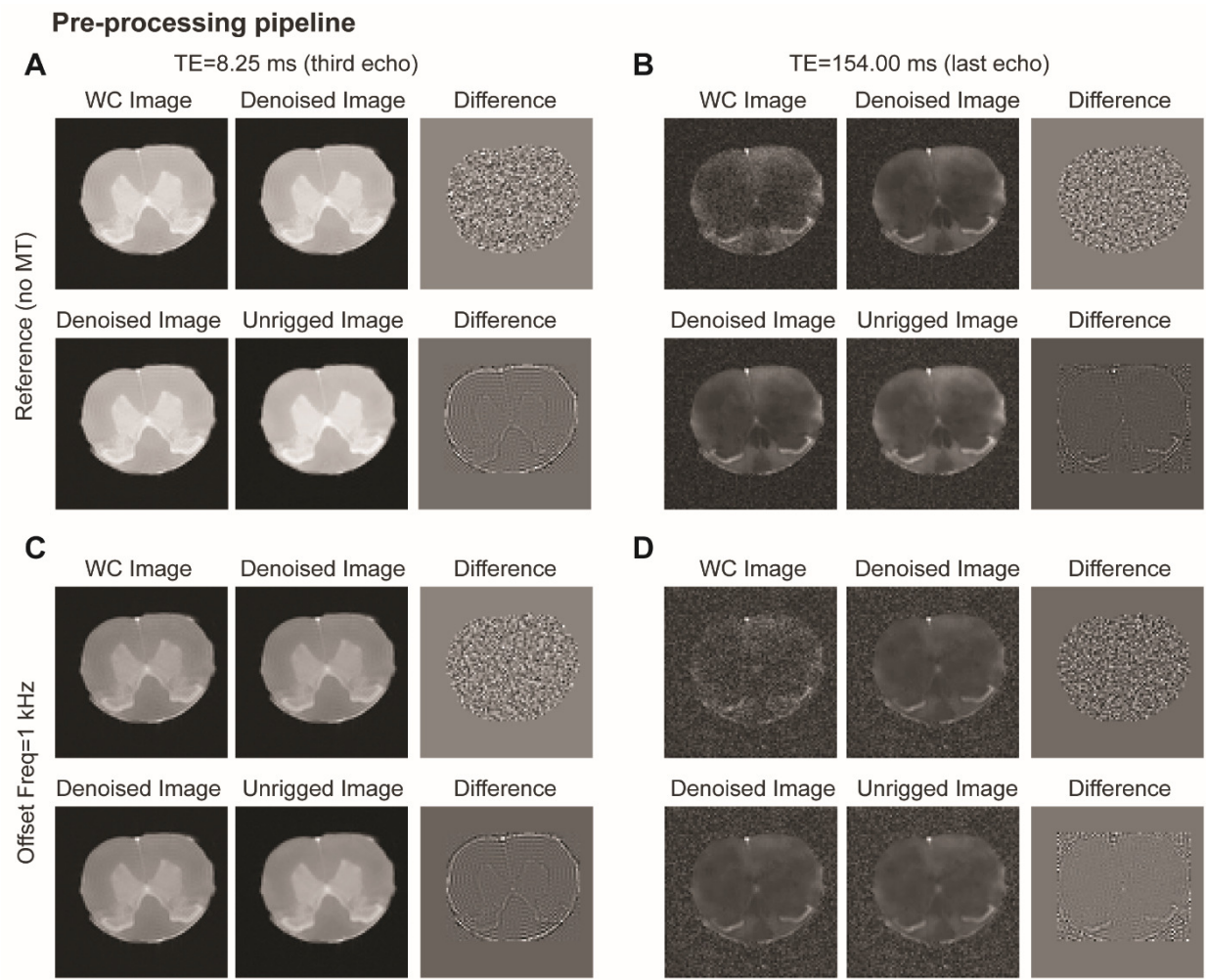

**Supplementary Figure 3 – Pre-processing pipeline and effects on a dataset (SC #1). A)** Magnitude data was converted to real data (Eichner et al., 2016) (top left). PCA denoising as described in (Veraart et al., 2016) is then applied (top centre). The difference map between denoised and Wald-corrected only images (top right) shows no structure. Finally, Gibbs unringing as described in Kellner et al. (Kelnner et al., 2015) is used (bottom centre). The effects of unringing are shown in the difference map (bottom right). Denoising effects in images with different intensities – with/without MT modulation and at TE$_3$/TE$_{40}$, **A /B)** and **C/D)** respectively, are shown.

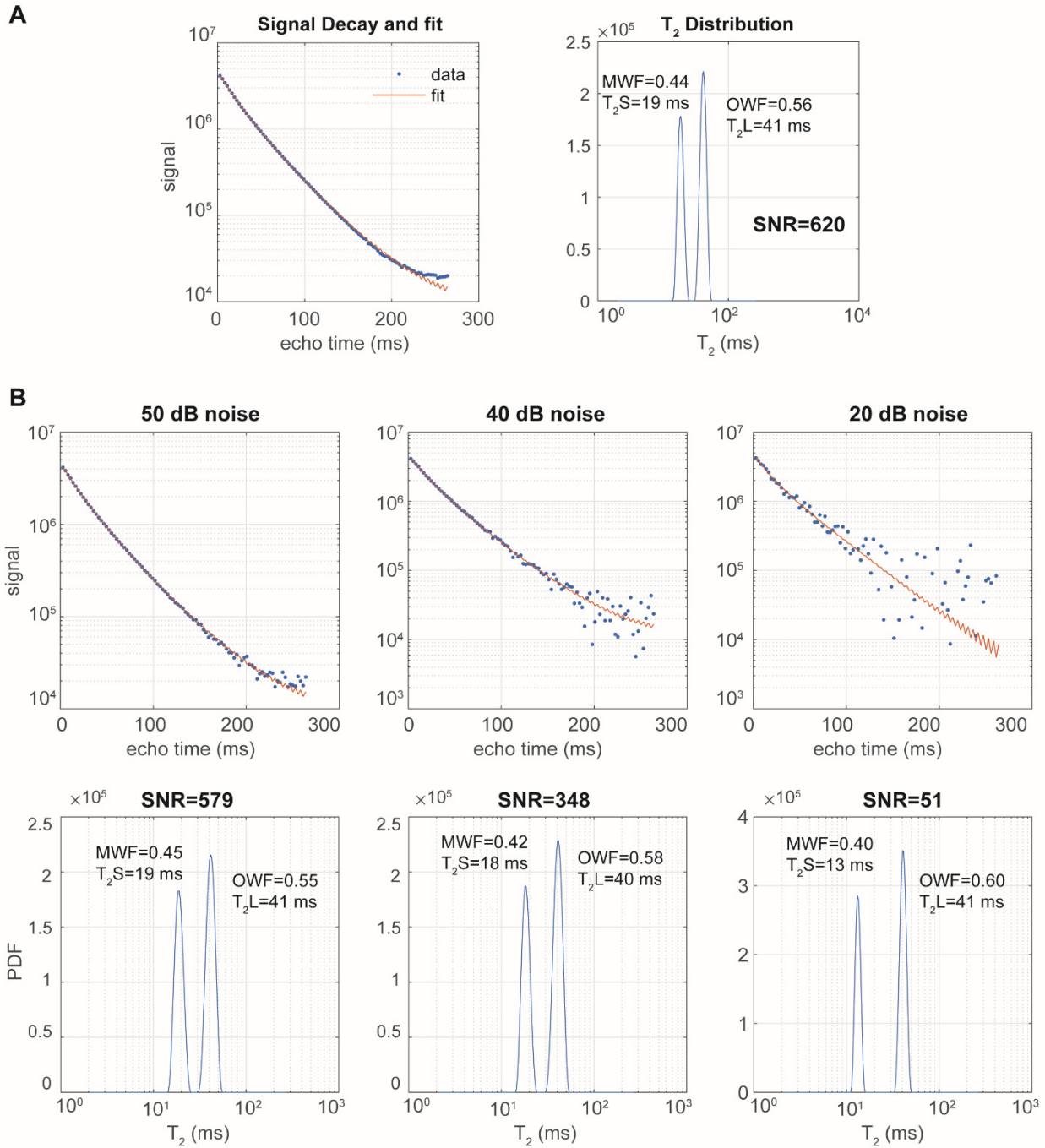

**Supplementary Figure 4 – Effects of adding gaussian noise to the signal decay in $T_2$ distributions. A)** Original $T_2$ decay and respective $T_2$ distribution. The SNR of the distribution (SNR=620) is given. The main characteristics of each peak (MWF/$T_2$S, OWF/$T_2$L) are reported next to the corresponding peak. B) Gaussian noise has added to the original plot. The amount of added noise was increase from left to right (from 50 dB to 20 dB). The resultant fit and respective $T_2$ distribution is shown per each noise level. The ability to estimate the peaks properties decreases with increasing noise level.

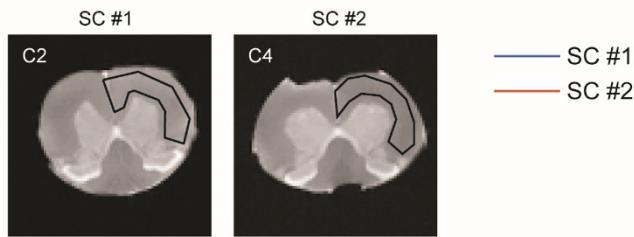
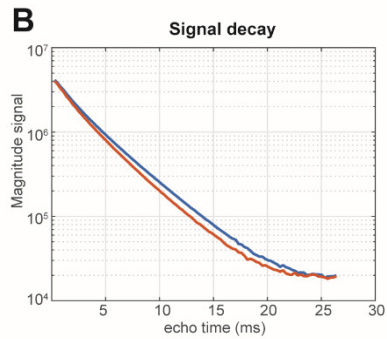
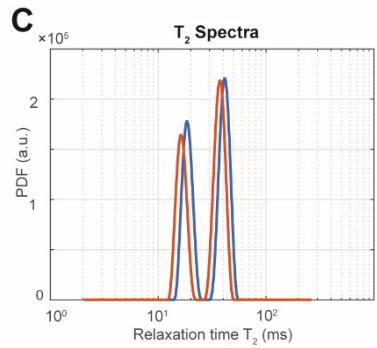
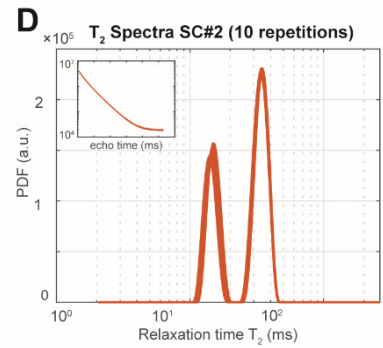
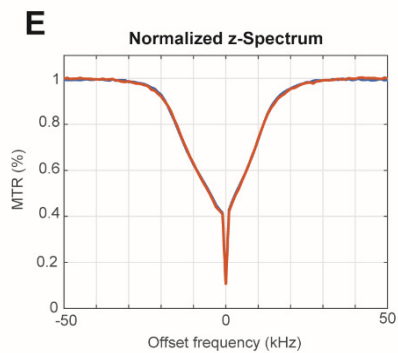
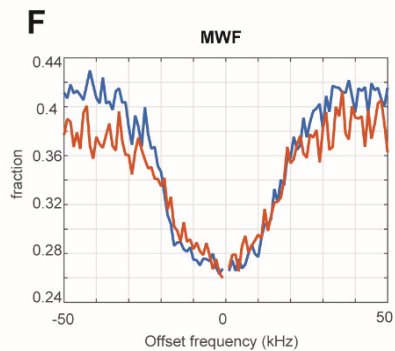
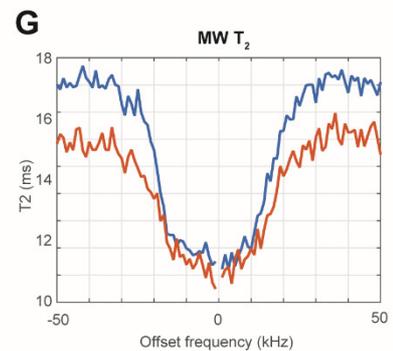

**Supplementary Figure 5 – Reproducibility: difference of result between two spinal cords.** The reproducibility of the experiment was tested. A second spinal cord was scanned in the same conditions as SC#1. **A)** The WM ROIs drawn for each SC are illustrated. **B)** Signal decay and **C)** $T_2$ spectra in the ROI in A) of the two spinal cords are plotted together. Variability in the decay is translated to slight shifts in the position of peaks in the $T_2$ distributions but the trend is kept. **D)** In order to ensure that the variability was not caused by acquisition errors, 10 repetitions of $T_2$ spectrum of SC#2 were acquired. The $T_2$ decays (inset) and $T_2$ spectra of the same sample are highly reproducible. **E)** Next, the variability of MT effect among different samples was tested. An overlap of z-spectra of the two spinal cords is observed. Finally, the variability in osi-

shifts was examined. The osi-shifts of **F)** fraction and **G)** $T_2$ of peak are similar in both spinal cords. The variability in these plots can be explained by variability on the relaxometry properties of different samples.

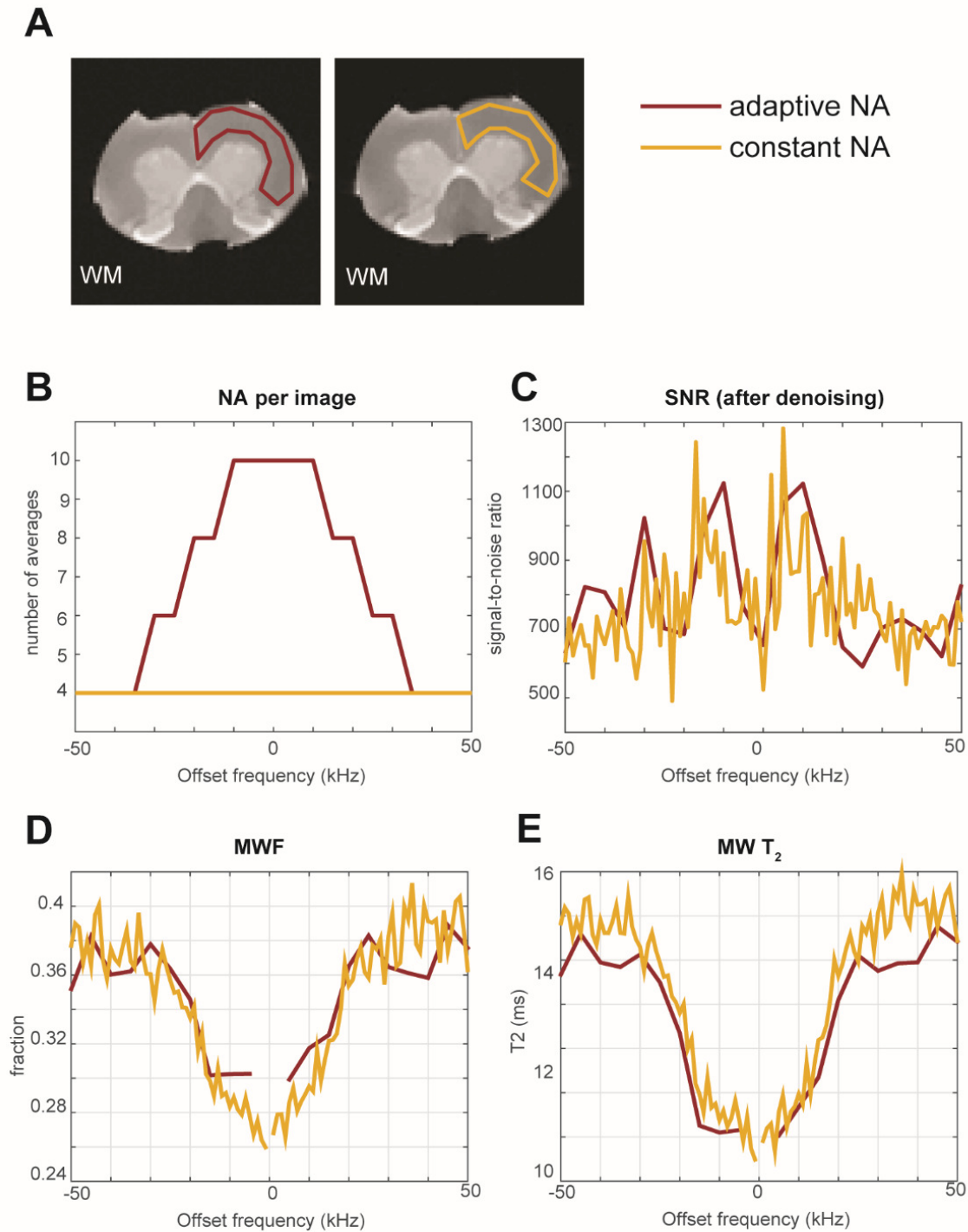

**Supplementary Figure 6 – Excluding SNR lost as source of osi-shift phenomena.**
A lower spectral resolution (offset frequencies=-50:5:50kHz) experiment was conducted in which the number of averages was progressively increased in scans using MT modulation closer to the water resonance (adaptive NA experiment). **A)** The different WM ROIs drawn for each experiment (constant vs adaptive NA) are shown (SC#2). **B)** The NA of each frequency point image is plotted for the adaptive and the constant NA experiments. **C)** The SNR was plotted against offset frequency irradiation. The SNR expected to be affected by MT modulation was brought back by increased efficiency of denoising due to a better characterisation of noise. This was observed in both experiments. As predicted, varying the NA (and consequently the SNR) did not modify the trends of both **D)** fraction and **E)** $T_2$ osi-shifts.

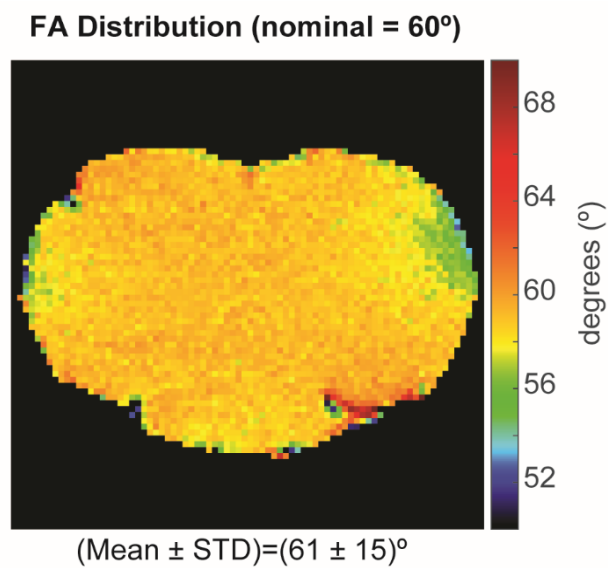

**Supplementary Figure 7 – Flip Angle (FA) Distribution with nominal FA=60º**. A double angle method (Stollberger et al., 1996) was applied with in an additional testing spinal cord with angles α=60º and 2α=120º for estimation of $B_1$ inhomogeneities. The Mean±STD FA over the entire SC was found to be (61±15)º. $B_1$ inhomogeneities were considered to be negligible.